\documentclass[journal]{IEEEtran}

\usepackage{amsmath,amsfonts}
\usepackage{amsmath}

\usepackage[table]{xcolor}
\usepackage{mathtools} 
\usepackage{array}
\usepackage{soul}
\usepackage{tabularray}
\usepackage{booktabs}
\usepackage[caption=false,font=footnotesize]{subfig}
\usepackage{textcomp}
\newtheorem{remark}{Remark}
\newtheorem{proposition}{Proposition}
\usepackage{stfloats}
\usepackage{url}
\usepackage{verbatim}
\usepackage{graphicx}
\usepackage{stfloats}
\usepackage{cite}
\usepackage{comment}
\usepackage{xspace}
\usepackage{xcolor}
\usepackage{enumitem}
\usepackage{amsfonts}
\usepackage{nicematrix}
\usepackage{ulem}
\usepackage{amssymb}

\usepackage{float}
\usepackage{tabularx}
\usepackage{subcaption} 
\usepackage[printonlyused]{acronym}
\usepackage[acronym]{glossaries}
\usepackage[ruled,vlined,linesnumbered,noend]{algorithm2e}
\SetKwInput{KwInput}{Input}
\SetKwInput{KwOutput}{Output}
\SetKw{KwRet}{return}
\usepackage[ruled,vlined,linesnumbered,noend]{algorithm2e}
\newacro{ISAC}{integrated sensing and communication}
\newacro{3GPP}{3rd Generation Partnership Project}
\newacro{SNR}{signal-to-noise ratio}
\newacro{MIMO}{multiple-input multiple-output}
\newacro{UT}{user terminal}
\newacro{UE}{user equipment}
\newacro{JIC}{koint imaging and communication}
\newacro{DFIC}{dual-function imaging and communication}
\newacro{BS}{base station}
\newacro{eMBB}{enhanced mobile broadband}
\newacro{mMTC}{massive machine-type communication}
\newacro{URLLC}{ultra-reliable and low-latency communication}
\newacro{DoF}{degree of freedom}
\newacro{mmWave}{millimeter-wave}
\newacro{CSI}{channel state information}
\newacro{MSE}{mean square error}
\newacro{ET}{extended target}
\newacro{PT}{point target}
\newacro{NF}{near-field}
\newacro{FF}{far-field}
\newacro{RCS}{radar cross section}
\newacro{SINR}{signal-to-interference-plus-noise ratio}
\newacro{XLAA}{extra large antenna array}
\newacro{AoSA}{array of sub-array}
\newacro{mMIMO}{massive MIMO}
\newacro{PA}{power amplifier}
\newacro{EE}{energy efficiency}
\newacro{QoS}{quality-of-service}
\newacro{EM}{electromagnetic}
\newacro{ULA}{uniform linear array}
\newacro{DFRC}{dual-functional radar communication}
\newacro{AoA}{angle of arrival}
\newacro{AoD}{angle of departure}
\newacro{AWGN}{additive white Gaussian noise}
\newacro{CRB}{Cramér-Rao bound}
\newacro{CUs}{communication users}
\newacro{CU}{communication user}
\newacro{RF}{radio frequency}
\newacro {FIM}{Fisher information matrix}
\newacro{RRT}{refreshing rate time}
\newacro{SDPs}{semidefinite programs}

\begin{document}
\title{\huge Energy-Efficient Target-Aware Hybrid Beamforming for THz Near-Field ISAC with Sparse Connectivity}

\author{Nusaibah A. Alshorman, Chong Han,~\IEEEmembership{Senior Member,~IEEE}, and H\"{u}seyin Arslan,~\IEEEmembership{Fellow,~IEEE}
\thanks{N. Alshorman and H. Arslan are with the Department of Electrical and Electronics Engineering, Istanbul Medipol University, Istanbul, Turkey. E-mail: nusaibah.abusanad@std.medipol.edu.tr, huseyinarslan@medipol.edu.tr. Chong Han is with the Terahertz Wireless Communications (TWC) Laboratory and the Cooperative Medianet Innovation Center (CMIC), School
of Information Science and Electronic Engineering, Shanghai Jiao Tong
University, Shanghai 200240, China (e-mail: chong.han@sjtu.edu.cn).}}
\maketitle
\begin{abstract}
Integrated sensing and communication (ISAC) at terahertz (THz) frequencies enables ultra-high-resolution perception while facing a key limitation: highly directional THz beams cannot illuminate extended targets within a single beam. Conventional solutions rely on sequential beam scanning, reducing sensing accuracy and increasing energy consumption. Moreover, in conventional sparse-array, grating lobes are generally treated as undesirable artifacts that should be suppressed to avoid ambiguity and interference. In contrast, this paper adopts a reverse design philosophy by intentionally engineering sparsity-induced grating lobes as controllable auxiliary illumination beams for extended-target sensing. This paper exploits grating lobes and proposes a sparse-connected hybrid beamforming architecture that intentionally engineers and exploits grating lobes to enable single-shot, full-aperture illumination of extended targets while supporting multi-user downlink communication. A switch-controlled sparse RF network preserves the array aperture and generates a dominant main lobe with structured secondary lobes covering the entire target extent. A covariance-driven alternating-minimization framework jointly optimizes digital precoders, quantized phase shifters, and antenna–RF switching. Simulations at 140 GHz demonstrate near fully-digital Cramér–Rao sensing accuracy, competitive communication performance in low-rank THz channels, rapid convergence, and significant hardware and energy savings, establishing structured sparse connectivity as a scalable and energy-efficient solution for extended-target THz ISAC.
\end{abstract}

\section{Introduction}

\IEEEPARstart{T}{he} terahertz (THz) band has emerged as a key enabler for future communication systems due to its ultra-wide bandwidth and extremely small wavelength. Operating from approximately 0.1~THz to 10 THz, this spectrum offers large contiguous bandwidths and enables ultra-high data rates and fine spatial resolution \cite{10439221,11060909}. These properties make THz particularly attractive for Integrated Sensing and Communication (ISAC), where a unified waveform and hardware platform jointly perform environmental sensing and data transmission \cite{10422712,elbir2024terahertz,rafique2025terahertz}. However, THz propagation is characterized by path loss, molecular absorption, and sparse scattering environments \cite{shen2025hybrid,11153494}. To compensate for attenuation and achieve sufficient link reliability, massive and ultra-massive MIMO arrays are typically employed. Yet, fully digital beamforming becomes impractical in THz systems because each antenna requires a dedicated RF chain, resulting in excessive hardware complexity and power consumption. Hybrid beamforming architectures have therefore emerged as practical solutions for THz beamforming and ISAC systems \cite{yang2025sensing,elbir2023near,yan2022energy}. A fundamental characteristic of ultra-large antenna arrays operating at THz frequencies is the significant expansion of the radiative near-field region. Unlike the far-field regime, where electromagnetic waves can be approximated as planar, near-field propagation is inherently spherical in nature. This introduces a joint dependence on both range and angle, resulting in distance-dependent beam patterns and spatial non-stationarity across the array~\cite{liu2025nf}. These effects become even more pronounced at THz frequencies due to the extremely small wavelength relative to the large aperture size, which in turn leads to a substantial increase in the Rayleigh distance~\cite{zhang2025nf}. From an ISAC perspective, conventional angle-only beamforming is no longer sufficient, since different scattering centers on an extended target may lie at different range-angle coordinates and therefore must be resolved jointly in both domains.

While hybrid architectures address hardware scalability, THz-ISAC introduces a deeper sensing challenge. Owing to the extremely high spatial resolution provided by ultra-wide bandwidth and large apertures, practical objects can no longer be accurately modeled as single point reflectors. Instead, they must be treated as \emph{extended targets} composed of multiple spatially distributed scattering centers \cite{11220947}. Moreover, at THz frequencies, the radar cross section (RCS) exhibits strong angular and frequency dependence \cite{wang2023propagation,rafique2025rough}. As a result, the backscattered energy is unevenly distributed across range-angle cells, with dominant reflections appearing only at specific observation angles.
This behavior creates a fundamental mismatch with conventional narrow-beam THz beamforming strategies. Highly directive beams concentrate most of the transmit power within an extremely narrow main lobe, while side lobes are intentionally suppressed. As a result, if the extended target does not fully overlap with the main-lobe footprint, a significant portion of its scattering region is illuminated with substantially lower array gain. Since the reflected echo power scales with the incident beamforming gain, this non-uniform and reduced illumination decreases the received echo energy, thereby degrading the sensing signal-to-noise ratio (SNR) and estimation accuracy.

\par

Recent studies have investigated sparse-array and hybrid-beamforming architectures to reduce hardware complexity in high-frequency and near-field systems~\cite{li2025sparse,evmorfos2024generative,liu2024doa,10475421}. In particular, sparse MIMO has become an increasingly attractive architecture for ISAC systems. By using fewer RF chains and antenna elements, it still manages to significantly boost the effective aperture and improve spatial resolution~\cite{li2025sparse}. At the same time, near-field beam-pattern behavior in both angle and range has been carefully characterized for sparse linear and extended coprime configurations~\cite{11039147}. However, these foundational frameworks focus strictly on fixed array layouts where the geometry is predetermined by the hardware design. In these communication-focused, fixed-topology designs, grating lobes are viewed as a liability because they cause inter-user interference, create angular ambiguities, and waste energy in unintended directions. Furthermore, even when the range dimension is accounted for, such as through the beam-depth metrics evaluated in~\cite{11039147}, it is not jointly optimized with angle to adaptively illuminate multiple scatterers across an extended target. Consequently, the prevailing approach in the literature has been limited to suppressing, mitigating, or otherwise controlling these grating-lobe effects rather than exploiting them. In sensing-oriented sparse-array designs, grating-lobe effects are also commonly mitigated through virtual-array construction, coarray processing, or spatial smoothing techniques~\cite{abusanad2024joint,dorvash2025virtual}. Meanwhile, extended-target modeling in ISAC has been explored mainly in lower-frequency regimes, where the target is represented by distributed scatterers and their correlation properties~\cite{wang2024cramer,du2022integrated,wang2023beamforming,yao2025hybrid,wang2025deep,zhao2024joint}.

However, existing THz hybrid beamforming frameworks largely assume point-target models or do not explicitly account for the angularly varying RCS of extended targets. Recent measurement-based studies conducted within the 3GPP ISAC standardization framework demonstrate that practical targets such as autonomous aerial vehicles exhibit pronounced directional scattering, with RCS variations on the order of 15-20 dB across azimuth angles \cite{11159279}. Despite this, the potential of deliberately engineered grating lobes for structured and uniform target illumination remains largely unexplored in THz-ISAC systems. \par

More broadly, the strategy of exploiting rather than mitigating spatial or spectral hardware characteristics for wide-area coverage has recently emerged across different contexts. In pure radar systems, spatial grating lobes have been leveraged to scale target search efficiency without serial scanning~\cite{hu2023}. Similarly, wideband mmWave/THz ISAC architectures have exploited beam-split and beam-squint anomalies to accomplish simultaneous multi-angle coverage across subcarriers~\cite{xu2023},
\cite{zhang2025}. These frameworks, however, rely on frequency-domain dispersion across a fully-connected analog network and do not accommodate extended-target models or spatial-domain optimization.\par

More broadly, the strategy of exploiting rather than mitigating spatial or spectral hardware characteristics for wide-area coverage has recently emerged across different contexts. In pure radar systems, spatial grating lobes have been leveraged to scale target search efficiency without serial scanning~\cite{hu2023}. While wideband mmWave/THz ISAC systems have harnessed beam-split and beam-squint phenomena to achieve simultaneous multi-angle coverage across subcarriers~\cite{xu2023,zhang2025}. These approaches, however, rely on frequency-domain dispersion within fully-connected analog front-ends and do not address extended-target sensing or spatial-domain optimization. This highlights a fundamental mismatch with near-field THz ISAC: conventional arrays are designed to suppress grating lobes to avoid ambiguity and interference, but for extended targets, this philosophy becomes counterproductive. An extended target often spans multiple range–angle cells with spatially separated dominant scatterers, meaning a single ultra-narrow THz beam typically illuminates only a fraction of the target support. To overcome this limitation, our framework deliberately shapes sparsity-induced lobes as controlled auxiliary illumination beams. Leveraging a sparse-connected hybrid architecture, the proposed design generates structured multi-lobe patterns that simultaneously cover the target's dominant scattering regions in a single snapshot. This approach preserves the effective aperture, satisfies communication constraints, and significantly reduces RF-chain count and hardware power consumption.

To systematically realize this concept, we formulate a wideband covariance-domain optimization problem that jointly designs sensing and multi-user communication under per-subcarrier transmit power and SINR constraints. The proposed formulation enables a sparse-connected hybrid precoder to approximate the performance of an ideal fully digital THz-ISAC system while remaining hardware-feasible. An efficient alternating-minimization algorithm is developed, combining convex covariance updates, closed-form analog phase-shifter refinement, and greedy switch selection for sparse connectivity learning.

The main contributions of this paper are summarized as follows:
\begin{itemize}
\item \textbf{Sparse-connected hybrid architecture for engineered multi-lobe illumination:}
We propose a sparse-connected hybrid beamforming framework that leverages, rather than suppresses, sparsity-induced grating lobes to support extended-target THz illumination. By exploiting structured antenna--RF sparsity through a switch-based RF network with reduced connectivity, the proposed architecture generates multiple spatial lobes capable of covering distributed scattering centers while lowering hardware complexity and power consumption.

    \item \textbf{Wideband covariance-domain joint radar-communication optimization:}  
We formulate a frequency-selective covariance-domain optimization framework that jointly satisfies extended-target sensing objectives and multi-user communication SINR constraints under hybrid hardware limitations. 

    \item \textbf{Comprehensive THz extended-target evaluation:}  
    Simulations at 140~GHz demonstrate that the proposed architecture closely approaches fully digital Cramér–Rao bounds for range–angle estimation while maintaining robust downlink communication performance and achieving substantial energy savings.
\end{itemize}

The following content of the paper is organized as follows. First, Section II details the wideband ISAC system architecture, alongside its channel and signal models for communication and sensing. Next, Section III defines problem formulation. Subsequently, Section IV analyzes our findings, and Section V concludes the paper.

Throughout the paper, capital boldface letters denote matrices, and lowercase boldface letters indicate vectors. The notation $\mathbf{A} \succeq \mathbf{0}$ means that $\mathbf{A}$ is a positive semi-definite matrix. The Hermitian, trace, and transpose operators are denoted by $(.)^H$, $\text{tr}(.)$, and $(.)^T$  respectively. The $\operatorname{diag}(.)$ denotes a diagonal matrix, $\circ$ denotes the Hadamard product, and $\otimes$ denotes the Kronecker product. $(\cdot)^\dagger$ denotes the Moore-Penrose pseudo-inverse
\section{The ISAC Architecture}

\begin{figure}[!t]
    \centering
    \includegraphics[width=1.0\linewidth, height=6cm]{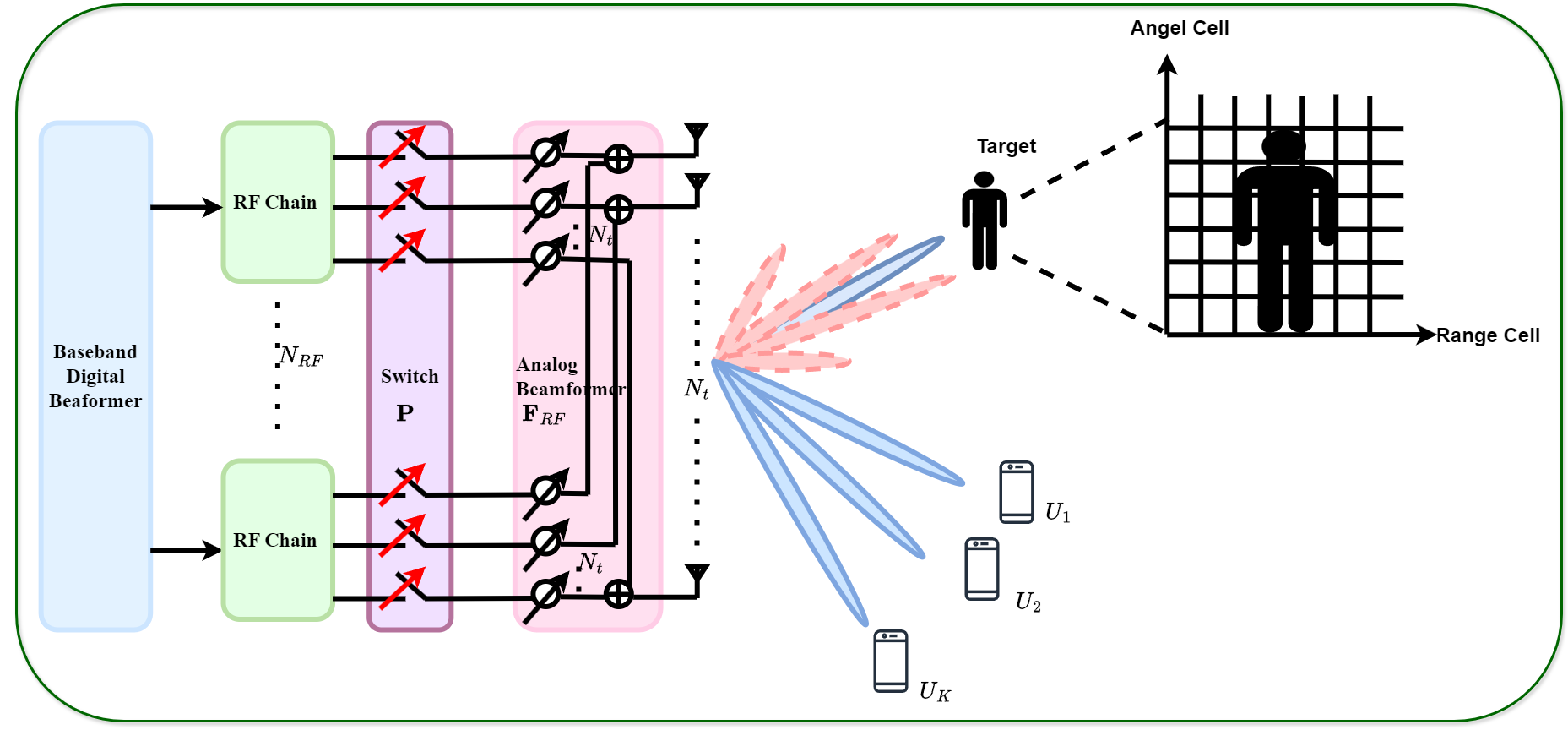} 
    \caption{System Model.}
    \label{system model}
\end{figure}

As illustrated in Fig. \ref{system model}, we consider a wideband massive-MIMO ISAC system, where the base station (BS) ISAC transmitter and radar receiver are integrated into a unified hardware platform. Equipped with $N_t$ transmit antennas and $N_r$ receive antennas with fully connected hybrid beamforming, the BS aims to serve $K$ \ac{CUs} and detect extended target simultaneously. 
The BS simultaneously transmits wireless signals serving two primary functions: (i) detecting targets located in the near-field region, and (ii) communicating with $K$ communication users (CUs), each equipped with its own ULA of $N_u = N_{t,u} = N_{r,u}$ antenna elements, denoted as $U_k$, where $k = 1, 2, \dots, K$. To meet the stringent requirements of simultaneous communication and sensing in THz integrated systems, extremely large bandwidths are essential to achieve both high data throughput and fine sensing resolution. Consequently, the conventional narrowband signal model commonly used at lower frequencies becomes inadequate for THz operation. In this work, we adopt an orthogonal frequency-division multiplexing (OFDM) waveform as a representative wideband signal model \cite{wang2024wideband}.


\subsection{OFDM Wideband Signal Model}

We consider a wideband THz ISAC system employing an orthogonal 
frequency-division multiplexing (OFDM) waveform. 
The total signal bandwidth $B$ is divided into $N_f$ subcarriers 
indexed by $n = 1,2,\ldots,N_f$, each centered at
\begin{equation}
    f_n = f_0 + (n-1)\Delta f,
\end{equation}
where $f_0$ denotes the lowest subcarrier frequency and 
$\Delta f = B/N_f$ is the subcarrier spacing. Since $\Delta f \ll f_c$, each subcarrier satisfies the narrowband
assumption. However, the variation of $f_n$ across tones introduces
frequency-dependent steering vectors and propagation phases, which
capture wideband near-field effects and beam-split phenomena inherent
to THz systems.

Let $L$ denote the number of OFDM symbols used for sensing and
communication processing. The joint sensing–communication symbol
matrix is defined as
\begin{equation}
    \mathbf{S} =
    \begin{bmatrix}
        \mathbf{S}_C \\[2pt]
        \mathbf{S}_R
    \end{bmatrix}
    \in \mathbb{C}^{N_s \times L},
    \qquad
    \frac{1}{L}\mathbf{S}\mathbf{S}^H = \mathbf{I},
    \label{eq:S_joint}
\end{equation}
where $\mathbf{S}_C$ and $\mathbf{S}_R$ denote the communication and
radar symbols, respectively, and
$\mathbb{E}[\mathbf{S}_C \mathbf{S}_R^H] = \mathbf{0}$.

At subcarrier $n$, the transmitted signal is
\begin{equation}
    \mathbf{X}[n] = \mathbf{F}[n] \mathbf{S}[n],
\end{equation}
where $\mathbf{F}[n] \in \mathbb{C}^{N_t \times N_s}$ denotes the
fully-digital or hybrid precoder.


\subsection{Array Geometry and Near-Field Steering Model}

The base station (BS) is equipped with a uniform linear array (ULA)
of $N_t$ antennas with inter-element spacing $d = \lambda_c/2$.
The antenna positions are
\begin{equation}
    x_m = \left(m - \frac{N_t - 1}{2}\right)d,
    \quad m=0,\ldots,N_t-1.
\end{equation}

For a point located at polar coordinates $(r,\theta)$ relative to the
array center, the spherical-wave propagation distance from antenna $m$ is
\begin{equation}
    r_m(r,\theta)
    =
    \sqrt{r^2 + x_m^2 - 2r x_m \sin\theta}.
    \label{eq:rm_nf}
\end{equation}

The frequency-dependent near-field steering vector at subcarrier $n$
is
\begin{equation}
    \big[\mathbf{a}_t(r,\theta;n)\big]_m
    =
    \frac{1}{\sqrt{N_t}}
    \exp\!\left(-j k_n r_m(r,\theta)\right),
    \label{eq:steer_nf_final}
\end{equation}
where $k_n = 2\pi f_n / c$.



\subsection{Wideband Downlink Communication Channel}

Each OFDM subcarrier is modeled as narrowband, while wideband
behavior is captured across tones via frequency-dependent steering.

Under a geometric channel model, the downlink channel between
the BS and communication user (CU) $k$ at subcarrier $n$ is
\begin{equation}
    \mathbf{H}_k[n]
    =
    \sum_{l=1}^{L_{ck}}
    \beta_{k,l}[n]\,
    \mathbf{a}_{u,k}(r_{k,l},\theta_{k,l};n)
    \mathbf{a}_t^H(r_{k,l},\theta_{k,l};n),
    \label{eq:Hk_final}
\end{equation}
where $L_{ck}$ denotes the number of propagation paths and
$\beta_{k,l}[n]$ is the complex frequency-dependent path gain.

The received signal at CU $k$ is
\begin{equation}
    \mathbf{y}_k[n]
    =
    \mathbf{H}_k[n]\mathbf{X}[n]
    +
    \mathbf{z}_k[n],
\end{equation}
where $\mathbf{z}_k[n] \sim \mathcal{CN}(\mathbf{0},\sigma_C^2\mathbf{I})$.


\subsection{Wideband Monostatic Sensing Channel}

We consider a monostatic sensing architecture in which the
transmit and receive arrays are co-located.
The extended target is modeled as $M$ discrete scatterers
located at $(r_i,\theta_i)$.

The wideband near-field sensing channel at subcarrier $n$ is
\begin{equation}
    \mathbf{G}[n]
    =
    \sum_{i=1}^{M}
    \alpha_i[n]\,
    \mathbf{a}_r(r_i,\theta_i;n)
    \mathbf{a}_t^H(r_i,\theta_i;n),
    \label{eq:G_final}
\end{equation}
where under the monostatic assumption
\begin{equation}
    \mathbf{a}_r(r_i,\theta_i;n)
    =
    \mathbf{a}_t(r_i,\theta_i;n).
\end{equation}

\subsubsection*{Scattering Coefficient Model}

The complex scattering coefficient is
\begin{equation}
    \alpha_i[n]
    =
    \gamma_i[n]\,
    e^{-j 2 k_n r_i}
    e^{j\phi_i},
    \label{eq:alpha_final}
\end{equation}
where $\phi_i \sim \mathcal{U}[0,2\pi)$ models random surface scattering,
and $e^{-j2k_n r_i}$ represents the two-way propagation phase.

The amplitude term follows the monostatic radar equation:
\begin{equation}
    \gamma_i[n]
    =
    \sqrt{
    \frac{
        \sigma_i(\theta_i,f_n)\,\lambda_n^2
    }{
        (4\pi)^3 r_i^4
    }},
    \label{eq:gamma_i_multi}
\end{equation}
where $\sigma_i(\theta_i,f_n)$ denotes the instantaneous radar cross-section (RCS) magnitude. 

\subsection{Transmit Architectures}
At subcarrier $n$, the transmitted signal is expressed as
\begin{equation}
\mathbf{X}[n] = \mathbf{W}_{\mathrm{TX}}[n]\mathbf{S},
\end{equation}
where $\mathbf{W}_{\mathrm{TX}}[n]$ depends on the adopted transmit architecture. We consider three architectures: fully digital (FD), hybrid fully connected (HFC), and the proposed sparse-connected (SC).

\subsubsection{Sparse-Connected Beamforming (SC)}
To introduce controllable spatial sparsity and enable structured multi-lobe illumination, we model the analog precoder as

\begin{equation}
    \mathbf{F}_{\mathrm{RF}} = \mathbf{P} \circ \mathbf{E},
    \label{eq:FRF_SPS}
\end{equation}
where $\mathbf{P} \in \{0,1\}^{N_t\times N_{\mathrm{RF}}}$ 
is the binary switch matrix indicating which antenna-RF connections are active, 
and $\mathbf{E}\in\mathbb{C}^{N_t\times N_{\mathrm{RF}}}$  contains constant-modulus phase-shifter coefficients satisfying
\begin{equation}
    |[\mathbf{E}]_{i,j}| = \alpha_0, \quad 
    \angle[\mathbf{E}]_{i,j} \in \mathcal{Q},
    \label{eq:E_phase}
\end{equation}
where $\alpha_0$ is the constant amplitude ( $\alpha_0=1/\sqrt{N_t}$), 
and $\mathcal{Q}$ is the set of feasible phase values. Therefore, the resulting transmit covariance is
\begin{equation}
\mathbf{R}_X^{\mathrm{SC}}[n]
=
(\mathbf{P}\circ\mathbf{E})
\mathbf{R}_{\mathrm{BB}}[n]
(\mathbf{P}\circ\mathbf{E})^H.
\label{eq:RXSC}
\end{equation}

Let $L_{\mathrm{SW}}$ denote the effective insertion loss of an active antenna-RF switch path. Since all active links are assumed to employ the same switch technology, the insertion loss is modeled as a fixed path-dependent factor \cite{garcia2017reduced,sobolewski2022state}.  Realizing this covariance in the presence of switch insertion loss requires the following pre-switch RF power at subcarrier $n$:
\begin{equation}
P_{\mathrm{RF,in}}^{\mathrm{SC}}[n] = \ell_{\mathrm{SW}} \operatorname{tr} \left( \mathbf{R}_{X}^{\mathrm{SC}}[n] \right).
\label{eq:Pin_SC}
\end{equation}

By explicitly learning the switch matrix $\mathbf{P}$ together with the digital covariance, the SC structure preserves effective aperture while significantly reducing RF-chain connectivity and hardware power consumption. 

\subsection{Communication Performance}

At subcarrier $n$, the received signal at communication user $k$ is
\begin{equation}
\mathbf{y}_k[n]
=
\mathbf{H}_k[n]\mathbf{X}[n]
+
\mathbf{z}_k[n],
\end{equation}
where $\mathbf{z}_k[n]\sim\mathcal{CN}(\mathbf{0},\sigma_C^2\mathbf{I})$. Let $\mathbf{w}_{r,k}[n]$ denote the linear receive combiner. The post-combiner signal is
\begin{equation}
r_k[n]
=
\mathbf{w}_{r,k}^H[n]\mathbf{H}_k[n]\mathbf{X}[n]
+
\mathbf{w}_{r,k}^H[n]\mathbf{z}_k[n].
\label{eq:rk_final}
\end{equation}

Using the transmit covariance decomposition
\begin{equation}
\mathbf{R}_X[n]
=
\mathbf{R}_C[n]
+
\mathbf{R}_R[n],
\label{eq:cov_decomp}
\end{equation}
where $\mathbf{R}_C[n]$ and $\mathbf{R}_R[n]$ denote the communication and radar transmit covariances, respectively, the SINR at user $k$ can be expressed in stream form as

\begin{equation}
\mathrm{SINR}_k[n]
=
\frac{
\big|
\mathbf{w}_{r,k}^H[n]
\mathbf{H}_k[n]
\mathbf{f}_{C,k}[n]
\big|^2
}{
I_k^{\mathrm{MUI}}[n]
+
I_k^{\mathrm{radar}}[n]
+
\sigma_C^2
\|\mathbf{w}_{r,k}[n]\|_2^2
},
\label{eq:SINR_stream_final}
\end{equation}
where
\begin{align}
I_k^{\mathrm{MUI}}[n]
&=
\sum_{j\neq k}
\big|
\mathbf{w}_{r,k}^H[n]
\mathbf{H}_k[n]
\mathbf{f}_{C,j}[n]
\big|^2,
\\[4pt]
I_k^{\mathrm{radar}}[n]
&=
\left\|
\mathbf{w}_{r,k}^H[n]
\mathbf{H}_k[n]
\mathbf{F}_R[n]
\right\|_2^2.
\end{align}

Equivalently, in covariance-domain form,
\begin{equation}
\small
\mathrm{SINR}_k[n]
=
\frac{
\mathbf{w}_{r,k}^H[n]
\mathbf{H}_k[n]
\mathbf{R}_{C,k}[n]
\mathbf{H}_k^H[n]
\mathbf{w}_{r,k}[n]
}{
\mathbf{w}_{r,k}^H[n]
\mathbf{H}_k[n]
\mathbf{R}_{\mathrm{int},k}[n]
\mathbf{H}_k^H[n]
\mathbf{w}_{r,k}[n]
+
\sigma_C^2
\|\mathbf{w}_{r,k}[n]\|_2^2
},
\label{eq:SINR_cov_final}
\end{equation}
where
\begin{equation}
\mathbf{R}_{\mathrm{int},k}[n]
=
\mathbf{R}_C[n]
-
\mathbf{R}_{C,k}[n]
+
\mathbf{R}_R[n],
\end{equation}
and $\mathbf{R}_{C,k}[n]$ denotes the covariance corresponding to the desired communication stream of user $k$.

\section{Problem Formulation}
We aim to design a wideband sparsely connected (SC) hybrid beamformer that (i) preserves multi-user downlink communication quality and (ii) provides reliable illumination of an extended target in the THz near-field. 

\subsection{SC Hybrid Design via Projection of a Fully-Digital Benchmark}
We first obtain a \emph{principled} fully-digital (FD) ISAC benchmark $\mathbf{W}_{\mathrm{TX}}^{\mathrm{FD}}[n]$ (derived in Sec.~\ref{sec:FD-benchmark}) that optimizes the sensing/communication trade-off under the same system model. The SC design is then posed as a \emph{hardware-feasible projection} of this FD solution onto the SC hybrid structure. Specifically, for each subcarrier $n$, the SC transmitted signal is
$
\mathbf X^{\mathrm{SC}}[n] = (\mathbf P \circ \mathbf E)\mathbf F_{\mathrm{BB}}[n]\mathbf S,
$
and the corresponding transmit covariance is
\begin{equation}
\mathbf R_X^{\mathrm{SC}}[n]
=
(\mathbf P \circ \mathbf E)\mathbf F_{\mathrm{BB}}[n]\mathbf F_{\mathrm{BB}}^{H}[n](\mathbf P \circ \mathbf E)^{H},
\label{eq:RX_SC_def_PF}
\end{equation}
where $\mathbf F_{\mathrm{BB}}[n]=[\mathbf F_{\mathrm{BB},C}[n],\,\mathbf F_{\mathrm{BB},R}[n]]$ collects the digital baseband precoders for the communication and sensing streams, respectively.

With these definitions, the SC hybrid beamforming problem is formulated as
\begin{subequations}
\label{eq:P1}
\begin{align}
\min_{\mathbf{P}, \mathbf{E}, \{\mathbf{F}_{\mathrm{BB}}[n]\}} \quad & \sum_{n=1}^{N_f} \Big\| \mathbf{W}_{\mathrm{TX}}^{\mathrm{FD}}[n]  - (\mathbf{P} \circ \mathbf{E}) \mathbf{F}_{\mathrm{BB}}[n] \Big\|_F^2 \label{eq:P1a} \\
\text{s.t.} \quad & \mathrm{SINR}_{k}[n]\left(\mathbf{R}_X^{\mathrm{SC}}[n]\right) \geq \gamma_k, \quad \forall k, n, \label{eq:P1b} \\
& \ell_{\mathrm{SW}} \mathrm{tr}\left(\mathbf{R}_X^{\mathrm{SC}}[n]\right) \leq P_{\max}, \quad \forall n, \label{eq:P1c} \\
& \mathrm{tr}(\mathbf{R}_X[n]\mathbf{B}_t(r_g, \theta_g; n)) \ge \eta_{\text{tgt}}, \quad \forall g \in \mathcal{G}, \, n, \label{eq:P1d} \\
& \mathrm{tr}\left(\mathbf{R}_X^{\mathrm{SC}}[n] \mathbf{B}_t(r_k, \theta_k; n)\right) \leq \rho_{\mathrm{CU}}, \quad \forall k, n, \label{eq:P1e} \\
& [\mathbf{P}]_{i,j} \in \{0, 1\}, \quad \forall i, j, \label{eq:P1f} \\
& \|\mathbf{P}_{:,j}\|_0 \le d_{\max}, \quad \forall j, \label{eq:P1g} \\
& [\mathbf{P}]_{1,j} = 1, \quad [\mathbf{P}]_{N_t,j} = 1, \quad \forall j, \label{eq:P1h} \\
& |[\mathbf{E}]_{i,j}| = \alpha_0, \quad \angle[\mathbf{E}]_{i,j} \in \mathcal{Q}, \quad \forall i, j. \label{eq:P1i}
\end{align}
\end{subequations}
where $
\mathbf B_t(r,\theta;n)
=
\mathbf a_t(r,\theta;n)
\mathbf a_t^H(r,\theta;n)
$ and $\|\mathbf a_t(r,\theta;n)\|_2=1$. \eqref{eq:P1b} enforces the per-subcarrier SINR requirements $\{\gamma_k\}$ for all communication users. This is the primary communication-quality constraint and prevents the sensing-oriented multi-lobe radiation from violating downlink QoS, \eqref{eq:P1c} imposes the per-subcarrier transmit power budget. 
\eqref{eq:P1d} ensures a minimum radiated energy toward \emph{every} grid point $(r_g,\theta_g)$, $g\in\mathcal G$, of a coarse region-of-interest (ROI) grid that covers the expected target occupancy region, rather than toward the (a priori unknown) true scatterer positions $(r_i,\theta_i)$ themselves. This constraint formalizes the requirement of \emph{single-shot} illumination.
Enforcing~\eqref{eq:P1d} across all $g\in\mathcal G$ and $n$ compels the transmit covariance to distribute energy over multiple angular directions in a frequency-consistent manner, which is crucial in the presence of wideband beam-split. 
Concretely, $\mathcal G=\{(r_g,\theta_g)\}_{g=1}^{G}$ is a coarse grid covering this ROI, with $G\ll M$ grid points typically sufficient. Constraint~\eqref{eq:P1e} is a \emph{spatial leakage control} constraint evaluated at the scheduled CU directions $(r_k,\theta_k)$. 
By bounding the radiated energy along the CU directions,~\eqref{eq:P1e} regularizes the beampattern and prevents pathological concentration of replica-lobe power toward users. The binary matrix $\mathbf P$ activates antenna-RF links via switches and~\eqref{eq:P1g} limits the fan-out of each RF chain to at most $d_{\max}$ active antenna connections. \eqref{eq:P1h} forces the first and last antennas to be active on every RF chain. 
Finally,~\eqref{eq:P1i} models practical phase-shifter hardware via constant-modulus and quantized-phase constraints. The amplitude $\alpha_0$ is fixed (e.g., $\alpha_0=1/\sqrt{N_t}$) and $\mathcal Q$ denotes the finite-resolution phase.

 Problem~\eqref{eq:P1} is nonconvex due to the binary variables $\mathbf P$, the constant-modulus/quantized phases in $\mathbf E$, and the bilinear coupling between analog and digital beamformers. We therefore adopt an alternating-minimization procedure that cyclically updates (i) $\{\mathbf F_{\mathrm{BB}}[n]\}$, (ii) $\mathbf E$, and (iii) $\mathbf P$ while fixing the other variables. 

\subsection{Fully-Digital Benchmark Optimization Problem Formulation}
\label{sec:FD-benchmark}

To establish a \emph{solid} and reproducible performance reference, the fully-digital (FD) benchmark
$\mathbf W_{\mathrm{TX}}^{\mathrm{FD}}[n]$ is obtained from a CRLB-aware wideband ISAC design, rather than being heuristically selected. 

\subsubsection{Wideband observation model}
Consider the monostatic radar receiver output at subcarrier $n$ over a frame of length $L$,
$\mathbf Y_R[n]\in\mathbb C^{N_r\times L}$, which follows the linear model
\begin{equation}
\mathbf Y_R[n]=\mathbf Y_0[n]+\mathbf Z_R[n],
\end{equation}
where $\mathbf Z_R[n]$ has i.i.d.\ entries distributed as $\mathcal{CN}(0,\sigma_R^2)$ and
\begin{equation}
\mathbf Y_0[n]=\sum_{i=1}^{M}\alpha_i[n]\mathbf A_i(r_i,\theta_i;n)\mathbf X[n].
\end{equation}
Stacking all subcarriers yields the global observation vector
\begin{equation}
\mathbf y
=
\begin{bmatrix}
\mathrm{vec}(\mathbf Y_R[1])\\[-2pt]
\vdots\\[-2pt]
\mathrm{vec}(\mathbf Y_R[N_f])
\end{bmatrix}
\sim
\mathcal{CN}\!\big(\boldsymbol\mu(\boldsymbol\Psi),\,\sigma_R^2\mathbf I\big),
\end{equation}
with mean
\begin{equation}
\boldsymbol\mu(\boldsymbol\Psi)
=
\begin{bmatrix}
\mathrm{vec}(\mathbf Y_0[1])\\[-2pt]
\vdots\\[-2pt]
\mathrm{vec}(\mathbf Y_0[N_f])
\end{bmatrix}
=
\sum_{i=1}^{M}\sum_{n=1}^{N_f}
\alpha_i[n]\,
\mathbf c_i[n],
\label{eq:mu_sum_multi}
\end{equation}
where
\begin{equation}
\mathbf c_i[n]
=
\mathrm{vec} \!\big(\mathbf A_i(r_i,\theta_i;n)\mathbf X[n]\big).
\label{eq:ci_def}
\end{equation}

\subsubsection{Parameterization and FIM structure}
We collect the \emph{geometric} parameters of the extended target as
\begin{equation}
\boldsymbol\Psi
=
\begin{bmatrix}
\boldsymbol\vartheta_1\\[-2pt]
\vdots\\[-2pt]
\boldsymbol\vartheta_M
\end{bmatrix}
\in\mathbb R^{2M},
\qquad
\boldsymbol\vartheta_i=[\,\theta_i,\ r_i\,]^T.
\label{eq:Psi_geometry}
\end{equation}
The per-scatterer random phases $\{\phi_i\}$ 
are treated as \emph{nuisance} terms.
With the Gaussian observation model above, the Fisher information matrix (FIM) for $\boldsymbol\Psi$ is
\begin{equation}
\mathbf J(\boldsymbol\Psi)
=
\frac{2}{\sigma_R^2}
\Re\!\left\{
\left(
\frac{\partial\boldsymbol\mu(\boldsymbol\Psi)}
     {\partial\boldsymbol\Psi^T}
\right)^{\!H}
\left(
\frac{\partial\boldsymbol\mu(\boldsymbol\Psi)}
     {\partial\boldsymbol\Psi^T}
\right)
\right\}
=
\frac{2}{\sigma_R^2}\Re\{\mathbf D^H\mathbf D\},
\label{eq:FIM_general}
\end{equation}
where $\mathbf D=\frac{\partial\boldsymbol\mu(\boldsymbol\Psi)}{\partial\boldsymbol\Psi^T}$.

\subsubsection{Effective derivatives and covariance-affine trace representation}
To capture the range/angle dependence of the amplitude factor $\gamma_i[n]$ in~\eqref{eq:gamma_i_multi}, we define
\begin{align}
\log \gamma_i[n]
&=
\frac{1}{2}\log\sigma_i(\theta_i,f_n)
+ \log\lambda_n
- \frac{3}{2}\log(4\pi)
- 2\log r_i,
\label{eq:log_gamma_multi}
\end{align}
and the normalized derivatives
\begin{equation}
d_{\theta_i}[n]=\frac{\partial}{\partial\theta_i}\log\gamma_i[n],
\qquad
d_{r_i}[n]=\frac{\partial}{\partial r_i}\log\gamma_i[n].
\label{eq:dtheta_dr_def_multi}
\end{equation}
Under the standard assumption that the RCS magnitude is range-invariant over the considered aperture and bandwidth,
$\partial\sigma_i(\theta_i,f_n)/\partial r_i=0$, we obtain $d_{r_i}[n]=-2/r_i$, while $d_{\theta_i}[n]$ retains angular dependence.

These lead to the effective derivative matrices
\begin{align}
\boldsymbol\Xi_{\theta_i}[n]
&=
d_{\theta_i}[n]\mathbf A_i(r_i,\theta_i;n)
+
\frac{\partial\mathbf A_i}{\partial\theta_i},
\label{eq:Xi_theta_final}
\\[1mm]
\boldsymbol\Xi_{r_i}[n]
&=
d_{r_i}[n]\mathbf A_i(r_i,\theta_i;n)
+
\frac{\partial\mathbf A_i}{\partial r_i}
-
j\frac{4\pi f_n}{c}\mathbf A_i(r_i,\theta_i;n),
\label{eq:Xi_r_final}
\end{align}

Hence, the Jacobian columns for $(\theta_i,r_i)$ can be written as
\begin{align}
\frac{\partial\boldsymbol\mu}{\partial\theta_i}
&=
\sum_{n=1}^{N_f}
\alpha_i[n]\,
\mathrm{vec}\!\big(\boldsymbol\Xi_{\theta_i}[n]\mathbf X[n]\big),
\label{eq:dmudtheta_multi}
\\
\frac{\partial\boldsymbol\mu}{\partial r_i}
&=
\sum_{n=1}^{N_f}
\alpha_i[n]\,
\mathrm{vec}\!\big(\boldsymbol\Xi_{r_i}[n]\mathbf X[n]\big).
\label{eq:dmudr_multi}
\end{align}

\emph{ (covariance-affine EFIM):}
Directly using~\eqref{eq:FIM_general} yields quadratic dependence on the precoders. To obtain a benchmark that is convex in transmit \emph{covariances}, we employ an equivalent-FIM (EFIM) formulation that preserves the dependence on $\mathbf X[n]$ through $\mathbf R_X^{\mathrm{FD}}[n]=\mathbb E[\mathbf X[n]\mathbf X^H[n]]$ and thus becomes affine in $\mathbf R_X^{\mathrm{FD}}[n]$.To see this explicitly, consider the angle-derivative column~\eqref{eq:dmudtheta_multi}. Using the identity $|\mathrm{vec}(\mathbf B\mathbf X[n])|^2=\mathrm{tr}(\mathbf B^H\mathbf B\,\mathbf X[n]\mathbf X^H[n])$ together with $\mathbb E[\mathbf X[n]\mathbf X^H[n]]=\mathbf R_X^{\mathrm{FD}}[n]$, the corresponding diagonal FIM contribution becomes
\begin{align}
\mathbb E\!\left[\Big|\frac{\partial\boldsymbol\mu}{\partial\theta_i}\Big|^2\right]
&=
\sum_{n=1}^{N_f}|\alpha_i[n]|^2\,\mathrm{tr}\!\big(\boldsymbol\Xi_{\theta_i}^H[n]\boldsymbol\Xi_{\theta_i}[n]\mathbf R_X^{\mathrm{FD}}[n]\big)
\\
&=
\sum_{n=1}^{N_f}|\gamma_i[n]|^2\,k_{\theta\theta}^{(i)}[n],
\label{eq:Etheta_deriv}
\end{align}
where  $|\alpha_i[n]|=|\gamma_i[n]|$ from~\eqref{eq:alpha_final}. The same steps applied to the range derivative~\eqref{eq:dmudr_multi} and to the cross term yield $k_{rr}^{(i)}[n]$ and $k_{\theta r}^{(i)}[n]$, respectively. Crucially, every entry of the FIM is reduced to a trace expression that is \emph{linear} in $\mathbf R_X^{\mathrm{FD}}[n]$, which is precisely what makes the resulting EFIM affine in the transmit covariance. Specifically, for each scatterer $i$ and subcarrier $n$, define the energy trace terms
\begin{align}
k_{\theta\theta}^{(i)}[n]
&=
\mathrm{tr}\!\big(
\boldsymbol\Xi_{\theta_i}[n]^H\,
\boldsymbol\Xi_{\theta_i}[n]\,
\mathbf R_X^{\text{FD}}[n]
\big),
\label{affine2}
\\
k_{rr}^{(i)}[n]
&=
\mathrm{tr}\!\big(
\boldsymbol\Xi_{r_i}[n]^H\,
\boldsymbol\Xi_{r_i}[n]\,
\mathbf R_X^{\text{FD}}[n]
\big),
\label{affine1}
\\
k_{\theta r}^{(i)}[n]
&=
\mathrm{tr}\!\big(
\boldsymbol\Xi_{\theta_i}[n]^H\,
\boldsymbol\Xi_{r_i}[n]\,
\mathbf R_X^{\text{FD}}[n]
\big),
\label{affine}
\end{align}
which are \emph{linear} in $\mathbf R_X^{\text{FD}}[n]$.

Aggregating across subcarriers with the deterministic amplitude weights $|\gamma_i[n]|^2$ gives
\begin{equation}
S_{pq}^{(i)}
=
\sum_{n=1}^{N_f}
|\gamma_i[n]|^2\,
k_{pq}^{(i)}[n],
\qquad
p,q\in\{\theta,r\}.
\label{eq:Spq_simplified}
\end{equation}

We then define the per-scatterer EFIM block for $(\theta_i,r_i)$ as
\begin{equation}
\mathbf J^{(i)}
=
\frac{2L}{\sigma_R^2}
\begin{bmatrix}
S_{\theta\theta}^{(i)} & \Re\{S_{\theta r}^{(i)}\}\\[2pt]
\Re\{S_{\theta r}^{(i)}\} & S_{rr}^{(i)}
\end{bmatrix}
\in\mathbb R^{2\times 2},
\qquad i=1,\dots,M,
\label{eq:Jeff_i_multi_theta_r_convex}
\end{equation}
which is affine in $\{\mathbf R_X^{\text{FD}}[n]\}_{n=1}^{N_f}$ and hence affine in
$\{\mathbf R_C^{\mathrm{FD}}[n],\mathbf R_R^{\mathrm{FD}}[n]\}$ via
$\mathbf R_X^{\text{FD}}[n]=\mathbf R_C^{\text{FD}}[n]+\mathbf R_R^{\text{FD}}[n]$.

The corresponding CRLBs for each scatterer are
\begin{equation}
\mathrm{CRB}(\theta_i)=\big[\mathbf J^{(i)^{-1}}\big]_{1,1},
\qquad
\mathrm{CRB}(r_i)=\big[\mathbf J^{(i)^{-1}}\big]_{2,2}.
\label{eq:CRLB_theta_r_i_convex}
\end{equation}

\subsubsection{Global information matrix and block structure}
The global information matrix for $\boldsymbol\Psi$ can be written in block form as
\begin{equation}
\small
\mathbf J
=
\begin{bmatrix}
\mathbf J_{11} & \mathbf J_{12} & \cdots & \mathbf J_{1M}\\
\mathbf J_{21} & \mathbf J_{22} & \cdots & \mathbf J_{2M}\\
\vdots         & \vdots         & \ddots & \vdots        \\
\mathbf J_{M1} & \mathbf J_{M2} & \cdots & \mathbf J_{MM}
\end{bmatrix}.
\label{global}
\end{equation}
When scatterer signatures are sufficiently separated (e.g., resolvable in range and/or angle under large bandwidth and aperture), the cross-terms $\mathbf J_{ij}$, $i\neq j$, become small compared to the diagonal blocks, (see Appendix~A), yielding the widely used approximation
\begin{equation}
\mathbf J \approx \mathrm{blkdiag}\big(\mathbf J^{(1)},\dots,\mathbf J^{(M)}\big).
\label{eq:J_blkdiag_final_convex}
\end{equation}

\subsubsection{CRLB-aware FD benchmark with SINR constraints}
Since $\mathbf J(\{\mathbf R_C^{\mathrm{FD}}[n]\},\{\mathbf R_R^{\mathrm{FD}}[n]\})$ is affine in the covariance variables through~\eqref{affine2}--\eqref{affine}, an E-optimal (worst-direction) sensing design admits the convex semidefinite representation
\begin{subequations}
\label{eq:optimization_problem}
\begin{align}
\max_{\eta,\;\{\mathbf{R}_C^{\mathrm{FD}}[n]\},\;\{\mathbf{R}_R^{\mathrm{FD}}[n]\}}
\quad & \eta
\\
\text{s.t.}\quad
& \mathbf{J}\!\left(\mathbf{R}_C^{\mathrm{FD}}[n],\mathbf{R}_R^{\mathrm{FD}}[n]\right)
- \eta\mathbf{I} \succeq \mathbf{0},
\label{C1_final}
\\
& \mathrm{SINR}_k[n] \ge \gamma_k,\qquad \forall k,n,
\label{C2_final}
\\
& \sum_{n=1}^{N_f}
\left(
\mathrm{tr}(\mathbf{R}_C^{\mathrm{FD}}[n])
+ \mathrm{tr}(\mathbf{R}_R^{\mathrm{FD}}[n])
\right)
\le P_t,
\label{C3_final}
\\
& \mathbf{R}_C^{\mathrm{FD}}[n] \succeq \mathbf{0},\quad
\mathbf{R}_R^{\mathrm{FD}}[n] \succeq \mathbf{0},\qquad \forall n.
\label{C4_final}
\end{align}
\end{subequations}
The SINR constraints are imposed directly in covariance form as
\begin{equation}
\label{eq:SINR_covariance}
\small
\begin{split}
&\mathrm{tr}\big(\mathbf{Q}_k[n]\mathbf{R}_{C,k}^{\mathrm{FD}}[n]\big) - \gamma_k \Big( \sum_{j\neq k}\mathrm{tr}\big(\mathbf{Q}_k[n]\mathbf{R}_{C,j}^{\mathrm{FD}}[n]\big) \\
&\quad + \mathrm{tr}\big(\mathbf{Q}_k[n]\mathbf{R}_R^{\mathrm{FD}}[n]\big) \Big) \ge \gamma_k \sigma_C^2 \|\mathbf{w}_{r,k}[n]\|_2^2
\end{split}
\end{equation}
where $\mathbf{R}_{C,k}^{\mathrm{FD}}[n]=\mathbf{f}_{C,k}[n]\mathbf{f}_{C,k}^H[n]$ and
$\mathbf{Q}_k[n]=\mathbf{H}_k^H[n]\mathbf{w}_{r,k}[n]\mathbf{w}_{r,k}^H[n]\mathbf{H}_k[n]$.
Problem~\eqref{eq:optimization_problem} is a convex SDP and can be efficiently solved with standard solvers (e.g., \textsc{CVX}), providing the FD benchmark covariances and the corresponding precoders $\mathbf W_{\mathrm{TX}}^{\mathrm{FD}}[n]$ used as the reference point for the proposed SC hybrid design.

 \subsection{Alternating-Minimization Algorithm}
\label{sec:altmin}

Problem~\eqref{eq:P1} is a mixed-integer, nonconvex program due to
(i)~the binary switch matrix $\mathbf P$,
(ii)~the constant-modulus and quantized-phase constraints on $\mathbf E$,
and (iii)~the bilinear coupling between the analog network
$\mathbf F_{\mathrm{RF}}=\mathbf P\circ\mathbf E$ and the digital precoders
$\{\mathbf F_{\mathrm{BB}}[n]\}$.
\paragraph*{Design principle and surrogate objective}
Since SINR and sensing constraints depend only on transmit covariance $\mathbf R_X^{\mathrm{SC}}[n]$, 
Step~1 optimizes digital variables in covariance domain.
Steps~2-3 then refine the hardware-constrained analog
structure (quantized phases and sparse connectivity) to reduce the precoder mismatch.
This yields a scalable algorithm whose expensive projection onto the feasible set
is carried out only in Step~1.

\subsubsection*{Step~1: Covariance-Domain Digital Update (fixed $\mathbf P,\mathbf E$)}
Let $\mathbf F_{\mathrm{RF}}=\mathbf P\circ\mathbf E$ be fixed. The SC transmit covariance
at subcarrier $n$ can be expressed as
\begin{equation}
\mathbf R_X^{\mathrm{SC}}[n]
=
\mathbf F_{\mathrm{RF}}\,\mathbf V[n]\,\mathbf F_{\mathrm{RF}}^{H},
\qquad
\mathbf V[n]= \mathbf F_{\mathrm{BB}}[n]\mathbf F_{\mathrm{BB}}^{H}[n],
\label{eq:V_definition_refined}
\end{equation}
where $\mathbf V[n]$ is the digital covariance in the RF-chain domain.
Directly optimizing $\mathbf F_{\mathrm{BB}}[n]$ is nonconvex because the constraints
\eqref{eq:P1b}--\eqref{eq:P1e} become quartic in the entries of $\mathbf F_{\mathrm{BB}}[n]$.
Instead, we optimize $\{\mathbf V[n]\}$, which renders all constraints affine.

We determine $\{\mathbf V[n]\}$ by minimizing a covariance-matching criterion
subject to the original SINR, power, sensing constraints:
\begin{subequations}
\label{eq:cov_subproblem_refined}
\begin{align}
    \min_{\mathbf V[n]} \quad &
    \sum_{n=1}^{N_f}
    \Big\|
        \mathbf R_X^{\text{FD}}[n]
        - \mathbf F_{\mathrm{RF}} \mathbf V[n] \mathbf F_{\mathrm{RF}}^H
    \Big\|_F^2,
    \label{eq:cov_obj}
    \\[3pt]
    \text{s.t.}\quad &
    \eqref{eq:P1b}, \eqref{eq:P1c}, \eqref{eq:P1d}, \eqref{eq:P1e}
    \\[2pt]
    &
    \mathbf V[n] \succeq \mathbf 0,
    \quad \forall\,n.
    \label{eq:cov_psd}
\end{align}
\end{subequations}

Since the objective \eqref{eq:cov_obj} is convex quadratic in $\mathbf V[n]$
and all constraints are affine in $\mathbf V[n]$, problem~\eqref{eq:cov_subproblem_refined}
is a convex conic program and can be solved efficiently (via \textsc{CVX}). 
\paragraph*{Recovering a digital precoder from $\mathbf V^\star[n]$}
After solving \eqref{eq:cov_subproblem_refined}, we obtain $\mathbf V^\star[n]\succeq\mathbf 0$.
Any factor $\mathbf F_{\mathrm{BB}}[n]$ satisfying
$\mathbf F_{\mathrm{BB}}[n]\mathbf F_{\mathrm{BB}}^H[n]=\mathbf V^\star[n]$
induces the \emph{same} $\mathbf R_X^{\mathrm{SC}}[n]$ 
A numerically stable choice is the eigenfactor
\begin{equation}
\mathbf V^\star[n]=\mathbf U_V[n]\boldsymbol\Lambda_V[n]\mathbf U_V^H[n],
\qquad
\mathbf F_{\mathrm{BB}}[n]=\mathbf U_V[n]\boldsymbol\Lambda_V^{1/2}[n],
\label{eq:FBB_recovery_refined}
\end{equation}
which uses up to $N_{\mathrm{RF}}$ streams. If a smaller stream budget $N_s<N_{\mathrm{RF}}$
is enforced, one may retain the $N_s$ dominant eigenmodes and, if needed,
apply standard covariance-fitting randomization; in our simulations we set
$N_s=N_{\mathrm{RF}}$, thus \eqref{eq:FBB_recovery_refined} is directly feasible.

\paragraph*{Feasibility guidance for $(\eta_{\mathrm{tgt}},\rho_{\mathrm{CU}})$}
\label{subsec:feasibility_refined}
To ensure feasibility, $\eta_{\mathrm{tgt}}$ and $\rho_{\mathrm{CU}}$ are selected 
conservatively via:
\begin{equation}
M\,\eta_{\mathrm{tgt}}\le P_{\max}.
\label{eq:eta_power_bound_refined}
\end{equation}
In practice, part of the power must satisfy the SINR constraints; thus we recommend choosing
$\eta_{\mathrm{tgt}}$ such that
\begin{equation}
M\,\eta_{\mathrm{tgt}} + P_{\mathrm{comm}}[n] \le P_{\max},
\label{eq:eta_comm_bound_refined}
\end{equation}
where $P_{\mathrm{comm}}[n]$ is obtained from a communication-only feasibility solve.
Moreover, the spatial cap $\rho_{\mathrm{CU}}$ must not contradict the SINR requirements.
A conservative necessary condition is
\begin{equation}
\rho_{\mathrm{CU}}\ge \max_k \frac{\gamma_k \sigma_C^2}{\|\mathbf h_k[n]\|^2},
\label{eq:rho_bound_refined}
\end{equation}
when $\mathbf h_k[n]$ is (approximately) aligned with $\mathbf a_t(r_k,\theta_k;n)$.
We use \eqref{eq:eta_comm_bound_refined}--\eqref{eq:rho_bound_refined} to select
$(\eta_{\mathrm{tgt}},\rho_{\mathrm{CU}})$ and avoid infeasible instances.

\subsection*{Step~2: Analog Phase-Shifter Update (Fixed $\mathbf P, \mathbf F_{\mathrm{BB}}[n]$)}
\label{subsec:step2_E}

With $\mathbf P$ and $\{\mathbf F_{\mathrm{BB}}[n]\}$ fixed, we update $\mathbf E$
to reduce the precoder mismatch in \eqref{eq:P1a} under constant-modulus and
$B$-bit phase quantization.
Let $\mathbf f_{\mathrm{RF},j}=\mathbf p_j\circ \mathbf e_j$ be the $j$th column of
$\mathbf F_{\mathrm{RF}}$.
Define the residual that excludes RF chain $j$
\begin{equation}
\mathbf R_j[n] =
\mathbf W_{\mathrm{TX}}^{\mathrm{FD}}[n] -
\sum_{\ell\neq j}\mathbf f_{\mathrm{RF},\ell}\,[\mathbf F_{\mathrm{BB}}[n]]_{\ell,:},
\label{eq:residual_step2}
\end{equation}

so that the dependence on $\mathbf f_{\mathrm{RF},j}$ is isolated.
For each active connection $(i,j)$ with $[\mathbf P]_{i,j}=1$, the
unconstrained LS update is obtained by minimizing
$\sum_n \|\mathbf R_j[n] - \mathbf f_{\mathrm{RF},j}[\mathbf F_{\mathrm{BB}}[n]]_{j,:}\|_F^2$
with respect to $[\mathbf f_{\mathrm{RF},j}]_i$, yielding
\begin{equation}
\tilde E_{i,j} =
\frac{\sum_{n=1}^{N_f} \left([\mathbf R_j[n]]_{i,:}\,[\mathbf F_{\mathrm{BB}}[n]]_{j,:}^H\right)}
{\sum_{n=1}^{N_f}\|[\mathbf F_{\mathrm{BB}}[n]]_{j,:}\|_2^2},
\qquad [\mathbf P]_{i,j}=1,
\label{eq:E_LS_refined}
\end{equation}
and $\tilde E_{i,j}=0$ if $[\mathbf P]_{i,j}=0$. We then project onto the hardware-feasible set:
(i) constant modulus $|E_{i,j}|=\alpha_0$ and (ii) $B$-bit quantization
$\angle E_{i,j}\in\mathcal Q$:
\begin{equation}
\begin{aligned}
    E_{i,j}^{(1)} &= \alpha_0 \frac{\tilde E_{i,j}}{|\tilde E_{i,j}|}, \\
    \angle E_{i,j} &= \arg\min_{q \in \mathcal{Q}} |\angle E_{i,j}^{(1)} - q|, \\
    E_{i,j} &= \alpha_0 e^{j \angle E_{i,j}}.
\end{aligned}
\label{eq:E_projection_refined}
\end{equation}
This coordinate-update + projection is low-complexity and directly enforces \eqref{eq:P1i}. The proposed algorithm is summarized in Algorithm \eqref{alg:step2_E_refined}.
 Phase resolution $B$ primarily affects convergence speed rather than final accuracy, as Step~1 re-optimizes digital covariance for any $\mathbf F_{\mathrm{RF}}$. At low $B$ the quantization cells of $\mathcal Q$ are wide, so the projection step in~\eqref{eq:E_projection_refined} can repeatedly flip between adjacent quantization levels across consecutive outer iterations before settling, requiring more iterations to reach the same stationary point that finer resolutions reach almost immediately.

\begin{algorithm}[t]
\footnotesize
\caption{Step 2: Analog Phase-Shifter Update}
\label{alg:step2_E_refined}
\SetKwInput{KwRequire}{Req.}
\SetKwInput{KwOutput}{Out.}
\KwRequire{$\mathbf{P}$, $\{\mathbf{F}_{\mathrm{BB}}[n]\}_{n=1}^{N_f}$, $\{\mathbf{W}_{\mathrm{TX}}^{\mathrm{FD}}[n]\}_{n=1}^{N_f}$, $\alpha_0$, $\mathcal{Q}$;} 
\KwOutput{$\mathbf{E}$.}

\BlankLine

Set $E_{i,j}=0$ for all $(i,j)$ with $P_{i,j}=0$\;

\For{$j=1$ \KwTo $N_{\mathrm{RF}}$}{
  \For{$n=1$ \KwTo $N_f$}{
    $\mathbf{R}_j[n]\leftarrow \mathbf{W}_{\mathrm{TX}}^{\mathrm{FD}}[n] -\sum_{\ell\neq j} (\mathbf{p}_\ell\circ \mathbf{e}_\ell)\,[\mathbf{F}_{\mathrm{BB}}[n]]_{\ell,:}$\;
  }

  \For{each $i$ with $P_{i,j}=1$}{
    $\displaystyle \tilde{E}_{i,j}\leftarrow \sum_{n=1}^{N_f} \big([\mathbf{R}_j[n]]_{i,:}\,[\mathbf{F}_{\mathrm{BB}}[n]]_{j,:}^H\big) \Big/ \sum_{n=1}^{N_f}\|[\mathbf{F}_{\mathrm{BB}}[n]]_{j,:}\|_2^2 $\;
    
    $E_{i,j}^{(1)}\leftarrow \alpha_0\,\tilde{E}_{i,j}/|\tilde{E}_{i,j}|$\;
    
    $\theta_{i,j}\leftarrow \arg\min_{q\in\mathcal{Q}}|\angle E_{i,j}^{(1)}-q|$ and $E_{i,j}\leftarrow \alpha_0 e^{j\theta_{i,j}}$\; 
  }
}
\end{algorithm}

\subsubsection*{Step~3: Sparse Switch-Pattern Update (fixed $\mathbf E,\{\mathbf F_{\mathrm{BB}}[n]\}$)}
\label{subsec:step3_refined}
Finally, we update $\mathbf P$ under the degree constraint $\|\mathbf P_{:,j}\|_0\le d_{\max}$
and the edge activation \eqref{eq:P1h}.
Since optimizing $\mathbf P$ is combinatorial, we adopt a greedy RF-chain-wise procedure.
For each RF chain $j$, we initialize the active set with the enforced edges
$S_j^{(0)}=\{1,N_t\}$ and then add antennas until $|S_j|=d_{\max}$.

To select the next antenna, we evaluate the \emph{incremental precoder mismatch}
caused by activating candidate index $i\notin S_j$:
\begin{equation}
\begin{split}
\Delta_{i,j} =\sum_{n=1}^{N_f} \Big( 
& \left\| \mathbf{W}_{\mathrm{TX}}^{\mathrm{FD}}[n] - \mathbf{F}_{\mathrm{RF}}^{(+i,j)} \mathbf{F}_{\mathrm{BB}}[n] \right\|_F^2 \\
- & \left\| \mathbf{W}_{\mathrm{TX}}^{\mathrm{FD}}[n] - \mathbf{F}_{\mathrm{RF}}^{(S_j)} \mathbf{F}_{\mathrm{BB}}[n] \right\|_F^2 \Big),
\end{split}
\label{eq:delta_switch_refined}
\end{equation}
where $\mathbf F_{\mathrm{RF}}^{(S_j)}$ denotes the current RF matrix and
$\mathbf F_{\mathrm{RF}}^{(+i,j)}$ is obtained by additionally setting $[\mathbf P]_{i,j}=1$.
We choose $i^\star=\arg\min_{i\notin S_j}\Delta_{i,j}$ and update
$S_j\leftarrow S_j\cup\{i^\star\}$.
If a minimum degree $d_{\min}$ is enforced, we continue additions until $|S_j|\ge d_{\min}$.  However, directly evaluating~\eqref{eq:delta_switch_refined} requires recomputing
$\mathbf F_{\mathrm{RF}}^{(+i,j)}\mathbf F_{\mathrm{BB}}[n]$
for every candidate and every subcarrier. This results in a computational cost of
$\mathcal O(N_tN_sN_f)$ per candidate, or
$\mathcal O(N_{rf}d_{max}N_t^2N_sN_f)$ overall, which scales quadratically with
$N_t$. This cost can be significantly reduced by noting that activating the
connection $(i,j)$ changes only a single entry of
$\mathbf F_{\mathrm{RF}}$. Hence, the update is rank one and can be expressed as
\begin{equation}
\mathbf F_{\mathrm{RF}}^{(+i,j)}
=
\mathbf F_{\mathrm{RF}}^{(S_j)}
+
\mathbf e_i\,[\mathbf E]_{i,j}\,\mathbf e_j^{T},
\label{eq:rank1_update}
\end{equation}
where $\mathbf e_i\in\mathbb R^{N_t}$ and
$\mathbf e_j\in\mathbb R^{N_{rf}}$ denote the corresponding standard basis
vectors. Next, define the residual matrix together with the $j$th row of the
baseband precoder as
\begin{equation}
\mathbf R_j[n]
=
\mathbf W_{\mathrm{TX}}^{\mathrm{FD}}[n]
-
\mathbf F_{\mathrm{RF}}^{(S_j)}
\mathbf F_{\mathrm{BB}}[n],
\qquad
\mathbf f_{\mathrm{BB},j}[n]
=
[\mathbf F_{\mathrm{BB}}[n]]_{j,:},
\label{eq:rank1_residual}
\end{equation}
which allows~\eqref{eq:delta_switch_refined} to be written in the closed form
\begin{equation}
\Delta_{i,j}
=
\sum_{n=1}^{N_f}
\left(
-2\,\mathrm{Re}
\left\{
[\mathbf E]_{i,j}
[\mathbf g_j[n]]_i^*
\right\}
+
|[\mathbf E]_{i,j}|^2
\,
\|\mathbf f_{\mathrm{BB},j}[n]\|_2^2
\right),
\label{eq:delta_closed_form}
\end{equation}
where
$\mathbf g_j[n]=\mathbf R_j[n]\mathbf f_{\mathrm{BB},j}^{H}[n]$.
After selecting the optimal antenna index $i^\star$, the residual is updated as
\begin{equation}
\mathbf R_j[n]
\leftarrow
\mathbf R_j[n]
-
\mathbf e_{i^\star}
[\mathbf E]_{i^\star,j}
\mathbf f_{\mathrm{BB},j}[n],
\qquad
\forall n,
\label{eq:residual_update}
\end{equation}
thereby avoiding the need to recompute the hybrid precoder. As a result, the
overall complexity is reduced to
$\mathcal O(N_{rf}d_{max}N_tN_sN_f)$, which grows linearly with $N_t$. The
complete procedure is summarized in
Algorithm~\ref{alg:step3_P_refined}.

\begin{algorithm}[t]
\footnotesize
\caption{Step 3: Sparse Switch-Pattern Update}
\label{alg:step3_P_refined}
\SetKwInput{KwRequire}{Req.}
\SetKwInput{KwOutput}{Out.}
\KwRequire{$\mathbf{E}$, $\{\mathbf{F}_{\mathrm{BB}}[n]\}_{n=1}^{N_f}$, $\{\mathbf{W}_{\mathrm{TX}}^{\mathrm{FD}}[n]\}_{n=1}^{N_f}$, $(d_{\min},d_{\max})$;} 
\KwOutput{$\mathbf{P}$.}

\BlankLine

Initialize $\mathbf{P}\leftarrow \mathbf{0}$\;

\For{$j=1$ \KwTo $N_{\mathrm{RF}}$}{
  $S_j\leftarrow \{1,N_t\}$; \,\, Set $P_{1,j}=P_{N_t,j}=1$ and form $\mathbf{F}_{\mathrm{RF}}=\mathbf{P}\circ\mathbf{E}$\;
  Initialize residual $\mathbf{R}_j[n]\leftarrow \mathbf{W}_{\mathrm{TX}}^{\mathrm{FD}}[n]-\mathbf{F}_{\mathrm{RF}}\mathbf{F}_{\mathrm{BB}}[n],\,\forall n$\;
  
  \While{$|S_j|<d_{\max}$ \textbf{and} $\|\mathbf{P}\|_0<L_{\max}$}{
    \For{$n=1$ \KwTo $N_f$}{
      $\mathbf{g}_j[n]\leftarrow\mathbf{R}_j[n]\,\mathbf{f}_{\mathrm{BB},j}[n]^H$ \textbf{and} $c_j[n]\leftarrow\|\mathbf{f}_{\mathrm{BB},j}[n]\|_2^2$\;
    }
    \For{each $i\notin S_j$}{
      $\Delta_{i,j}\leftarrow\sum_{n=1}^{N_f}\big(-2\,\mathrm{Re}\{[\mathbf{E}]_{i,j}[\mathbf{g}_j[n]]_i^*\}+|[\mathbf{E}]_{i,j}|^2c_j[n]\big)$\;
    }
    Choose $i^\star=\arg\min_{i\notin S_j}\Delta_{i,j}$ \textbf{and} update $S_j\leftarrow S_j\cup\{i^\star\}$, $P_{i^\star,j}=1$\;
    Update $\mathbf{R}_j[n]\leftarrow\mathbf{R}_j[n]-\mathbf{e}_{i^\star}[\mathbf{E}]_{i^\star,j}\mathbf{f}_{\mathrm{BB},j}[n],\,\forall n$\;
  }
  
  \If{$|S_j|<d_{\min}$}{
    Add remaining indices via min $\Delta_{i,j}$ until $|S_j|=d_{\min}$\;
  }
}
\end{algorithm}

\paragraph*{Constraint handling across iterations.}
Step~2 and Step~3 aim to reduce the precoder mismatch under hardware constraints
but do not explicitly enforce \eqref{eq:P1b}--\eqref{eq:P1e} at intermediate iterates.
Global feasibility is re-established in the \emph{next} Step~1, which projects the design
back onto the set satisfying SINR, power, illumination, and cap constraints via
\eqref{eq:cov_subproblem_refined}. This separation keeps Steps~2--3 lightweight,
while Step~1 guarantees constraint compliance at every outer iteration after the projection. Finally, the overall Altmin algorithm is summarized in Algorithm \eqref{alg:ISAC_SC_altmin_main_refined}
\begin{algorithm}[t]
\footnotesize

\caption{AltMin for ISAC Sparse-Connected Hybrid Beamforming}
\label{alg:ISAC_SC_altmin_main_refined}
\SetKwInput{KwRequire}{Req.}
\SetKwInput{KwOutput}{Out.}
\KwRequire{$\{\mathbf{W}_{\mathrm{TX}}^{\mathrm{FD}}[n]\}$, $\{\mathbf{H}_k[n]\}$, $\{\gamma_k\}$, $\{\mathbf{B}_t\}$, $P_{\max}$, $\rho_{\mathrm{CU}}$, $(d_{\min},d_{\max})$, $(\alpha_0,\mathcal{Q})$, $(\varepsilon,t_{\max})$;} 
\KwOutput{$\mathbf{P}$, $\mathbf{E}$, and $\{\mathbf{F}_{\mathrm{BB}}[n]\}$.}

\BlankLine

\textbf{Initialization:}\\
Choose valid $\mathbf{P}^{(0)}$; Set active $|E_{i,j}^{(0)}|=\alpha_0$, $\angle E_{i,j}^{(0)}\in\mathcal{Q}$ (else $0$ if $P^{(0)}_{i,j}=0$)\;
$\mathbf{F}_{\mathrm{RF}}^{(0)}=\mathbf{P}^{(0)}\circ \mathbf{E}^{(0)}$ and $\mathbf{F}_{\mathrm{BB}}^{(0)}[n]=\arg\min_{\mathbf{F}_{\mathrm{BB}}[n]}\|\mathbf{W}_{\mathrm{TX}}^{\mathrm{FD}}[n]-\mathbf{F}_{\mathrm{RF}}^{(0)}\mathbf{F}_{\mathrm{BB}}[n]\|_F^2$\;
Compute initial mismatch metric $f^{(0)}$ and set $t=0$\;

\BlankLine

\While{$t<t_{\max}$}{
  $t\leftarrow t+1$ and set temporary matrix $\mathbf{F}_{\mathrm{RF}}^{(t-1)}=\mathbf{P}^{(t-1)}\circ \mathbf{E}^{(t-1)}$\;

  \tcp{Step 1: Digital Covariance Update \& Feasibility Projection}
  Solve convex program \eqref{eq:cov_subproblem_refined} with $\mathbf{F}_{\mathrm{RF}}^{(t-1)}$ to find $\{\mathbf{V}^\star[n]\succeq 0\}$\;
  Recover $\mathbf{F}_{\mathrm{BB}}^{(t)}[n]$ via EVD: $\mathbf{V}^\star[n]=\mathbf{U}_V[n]\boldsymbol\Lambda_V[n]\mathbf{U}_V^H[n]\rightarrow\mathbf{F}_{\mathrm{BB}}^{(t)}[n]=\mathbf{U}_V[n]\boldsymbol\Lambda_V^{1/2}[n]$\;

  \tcp{Step 2: Analog Phase Update}
  $\mathbf{E}^{(t)}\leftarrow$ Alg.~\ref{alg:step2_E_refined}$(\mathbf{P}^{(t-1)},\{\mathbf{F}_{\mathrm{BB}}^{(t)}[n]\},\{\mathbf{W}_{\mathrm{TX}}^{\mathrm{FD}}[n]\},\alpha_0,\mathcal{Q})$\;

  \tcp{Step 3: Switch-Pattern Update}
  $\mathbf{P}^{(t)}\leftarrow$ Alg.~\ref{alg:step3_P_refined}$(\mathbf{E}^{(t)},\{\mathbf{F}_{\mathrm{BB}}^{(t)}[n]\},\{\mathbf{W}_{\mathrm{TX}}^{\mathrm{FD}}[n]\},d_{\min},d_{\max},L_{\max})$\;

  \BlankLine
  Update $\mathbf{F}_{\mathrm{RF}}^{(t)}=\mathbf{P}^{(t)}\circ \mathbf{E}^{(t)}$ and calculate updated mismatch metric $f^{(t)}$\;
  
  \If{$\big|f^{(t)}-f^{(t-1)}\big|\,\big/\,f^{(t-1)}\le \varepsilon$}{
     \textbf{break}\;
  }
}
\end{algorithm}
\subsubsection*{Computational complexity}
The per-outer-iteration complexity is dominated by Step~1. The optimization variable per tone is $\mathbf V[n]\in\mathbb C^{N_{\mathrm{RF}}\times N_{\mathrm{RF}}}$,
hence the conic dimension scales with $N_{\mathrm{RF}}$ (not $N_t$), enabling large-array operation when $N_{\mathrm{RF}}\ll N_t$. 
Table~\ref{tab:complexity} summarizes the orders.

\begin{table}[h!]
\centering
\caption{Per-outer-iteration complexity and Numerical values in floating-point operations (FLOPs). $N_t=128$, $N_{rf}=8$, $N_f=64$, $N_s=8$, $d_{max}=24$, $K=4$, $M=6$, $N_{con}=K+M+1=11$.}
\label{tab:complexity}
\renewcommand{\arraystretch}{1.4}
\resizebox{\columnwidth}{!}{ 
\begin{tabular}{@{}llcc@{}}
\toprule
\textbf{Step} & \textbf{Operation} & \textbf{Complexity order} & \textbf{FLOPs}\\
\midrule
Step~1 & Covariance-domain SDP  & $\mathcal{O}(N_{rf}^6 + N_{con}N_{rf}^4)$ & $\approx 3.1\times10^5$\\
Step~1 & Wideband  & $\mathcal{O}\left(N_f(N_{rf}^6 + N_{con}N_{rf}^4)\right)$ & $\approx 2.0\times10^7$\\
Step~2 & Analog phase update & $\mathcal{O}(N_{rf},N_t,N_f,N_s)$ & $\approx 5.2\times10^5$\\
Step~3 & Greedy switch (rank-1 update) & $\mathcal{O}(N_{rf},d_{max},N_t,N_f,N_s)$ & $\approx 1.3\times10^7$\\
\midrule
 & \textbf{Total per outer iteration} & --- & $\approx 3.3\times10^7$\\
\bottomrule
\end{tabular}
}
\end{table}
\subsection{Power Consumption Model and Energy Efficiency Metric}


\subsubsection{Radiated Power}

For subcarrier $n$, the transmit covariance matrix is $\mathbf{R}[n] \in \mathbb{C}^{N_t \times N_t}$. The total radiated power across all subcarriers is $
P_{\mathrm{tx}}
=
\sum_{n=1}^{N_f} \mathrm{tr}\big(\mathbf{R}[n]\big).$

\subsubsection{Hardware Power Consumption}
The overall power consumption is modeled as

\begin{equation}
\begin{split}
P_{\mathrm{tot}}^{\mathrm{SC}} = \, &\ell_{\mathrm{SW}} P_{\mathrm{tx}}^{\mathrm{eff}} + P_{\mathrm{BB}} + N_{\mathrm{RF}} P_{\mathrm{RF}} \\
& + N_{\mathrm{PS}} P_{\mathrm{PS}} + \|\mathbf{P}\|_0 P_{\mathrm{SW}} + P_{\mathrm{reconf}}.
\end{split}
\label{eq:Ptotal}
\end{equation}
where $P_{\mathrm{tx}}^{\mathrm{eff}} =\frac{P_{\mathrm{tx}}}{\eta_{\mathrm{PA}}}$ is the  effective transmit power, $\eta_{\mathrm{PA}}\in(0,1]$ denotes the PA efficiency, $P_{\mathrm{BB}}$ denotes baseband processing power, $P_{\mathrm{RF}}$ is the power per RF chain, $P_{\mathrm{PS}}$ is the power per phase shifter, and $P_{\mathrm{SW}}$ is the power per active switch. The term $P_{\mathrm{reconf}}$ accounts for the power consumed during RF-chain activation and switch-state reconfiguration between consecutive sensing or beam-management updates. It is expressed as
\begin{equation}
P_{\mathrm{reconf}}
=
\frac{N_{\mathrm{RF}}E_{\mathrm{RF}} + N_{\mathrm{SW}}E_{\mathrm{SW}}}
{T_{\mathrm{upd}}},
\label{eq:Preconf}
\end{equation}
where $E_{\mathrm{RF}}$ and $E_{\mathrm{SW}}$ denote the energy required for RF-chain and switch-state transitions, respectively, and $T_{\mathrm{upd}}$ is the update interval. In the proposed architecture, the switch configuration remains unchanged throughout each sensing frame. As a result, no reconfiguration is performed during frame operation, and the associated power consumption is negligible under the quasi-static assumption considered in this work. Accordingly, $P_{\mathrm{reconf}}$ is set to zero unless otherwise specified~\cite{sobolewski2022state,ziegahn2025full}.

\subsubsection{Hardware Feasibility and Switch Non-Idealities}
\label{subsec:hw_feasibility}

 The relative energy-efficiency (EE) degradation due to an insertion loss of $L$~dB is
\begin{equation}
\frac{\mathrm{EE}^{(L)}}{\mathrm{EE}^{(0)}}
=
\frac{P_{\mathrm{circ}}+P_{\mathrm{tx}}^{\mathrm{eff}}}
{P_{\mathrm{circ}}+\ell_{\mathrm{SW}}P_{\mathrm{tx}}^{\mathrm{eff}}},
\label{eq:EE_loss_ratio}
\end{equation}
where $P_{\mathrm{circ}}$ is the total circuit power. Owing to its lower circuit complexity, SC degrades noticeably more slowly than FC as $L$ increases (Table~\ref{tab:EE_loss}); 
With $\|\mathbf P\|_0=192$ active links versus $N_tN_{\mathrm{RF}}=1024$ for FC, SC also achieves a $7.3$~dB ($18.75\%$) reduction in switching-related circuit overhead.

\begin{table}[htbp]
\centering
\caption{Energy efficiency relative to the ideal case versus insertion loss (SNR~$=15$~dB).}
\label{tab:EE_loss}
\renewcommand{\arraystretch}{1.25}
\footnotesize
\begin{tabular}{lccc}
\hline
Method & $L=0$~dB & $L=1$~dB & $L=3$~dB\\
\hline
SC-AltMin & $\mathrm{EE}_{\mathrm{SC}}^{(0)}$ & $0.89\times\mathrm{EE}_{\mathrm{SC}}^{(0)}$ & $0.50\times\mathrm{EE}_{\mathrm{SC}}^{(0)}$\\
FC Hybrid & $\mathrm{EE}_{\mathrm{FC}}^{(0)}$ & $0.74\times\mathrm{EE}_{\mathrm{FC}}^{(0)}$ & $0.32\times\mathrm{EE}_{\mathrm{FC}}^{(0)}$\\
\hline
\end{tabular}
\end{table}

Regarding switching speed, $\mathbf P$ is configured at the frame level and held fixed within each sensing--communication interval, so the relevant condition is $T_{\mathrm{SW}}\ll T_{\mathrm{upd}}$, with $T_{\mathrm{upd}}\approx10$--$100$~ms ~\cite{garcia2017reduced}. Representative THz-compatible switch technologies, FET/HEMT ($<1$~ns), RF MEMS ($1$--$100$~$\mu$s), and ferroelectric ($10$~ns--$1$~$\mu$s), all satisfy this requirement ~\cite{sobolewski2022state}, 
its only effect is the reconfiguration overhead $P_{\mathrm{reconf}}$ of~\eqref{eq:Preconf}, which is zero for static operation and remains below $400$~nW even in a worst-case dynamic scenario ($N_{\mathrm{SW}}^{\mathrm{tog}}\leq384$ toggles, $1$--$10$~pJ per toggle).

\section{Numerical Results}
\label{sec:numerical_results}
\begin{figure*}[!htbp]
\centering
\subfloat[Digital Beamforming (FD).]{%
\includegraphics[width=0.3\linewidth]{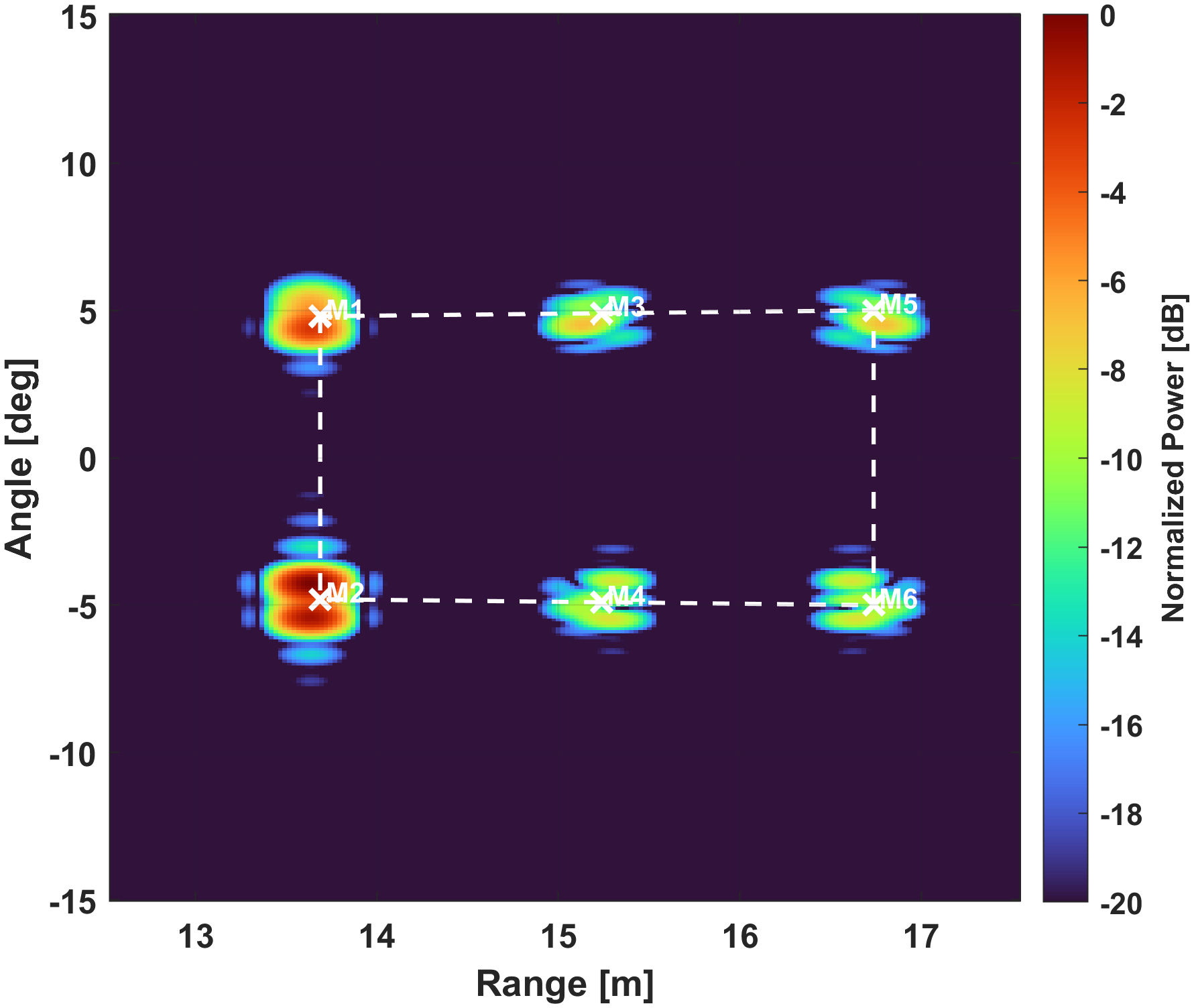}\label{fig:fd}}
\hfill
\subfloat[Full connected Hybrid Beamforming (FC).]{%
\includegraphics[width=0.3\linewidth]{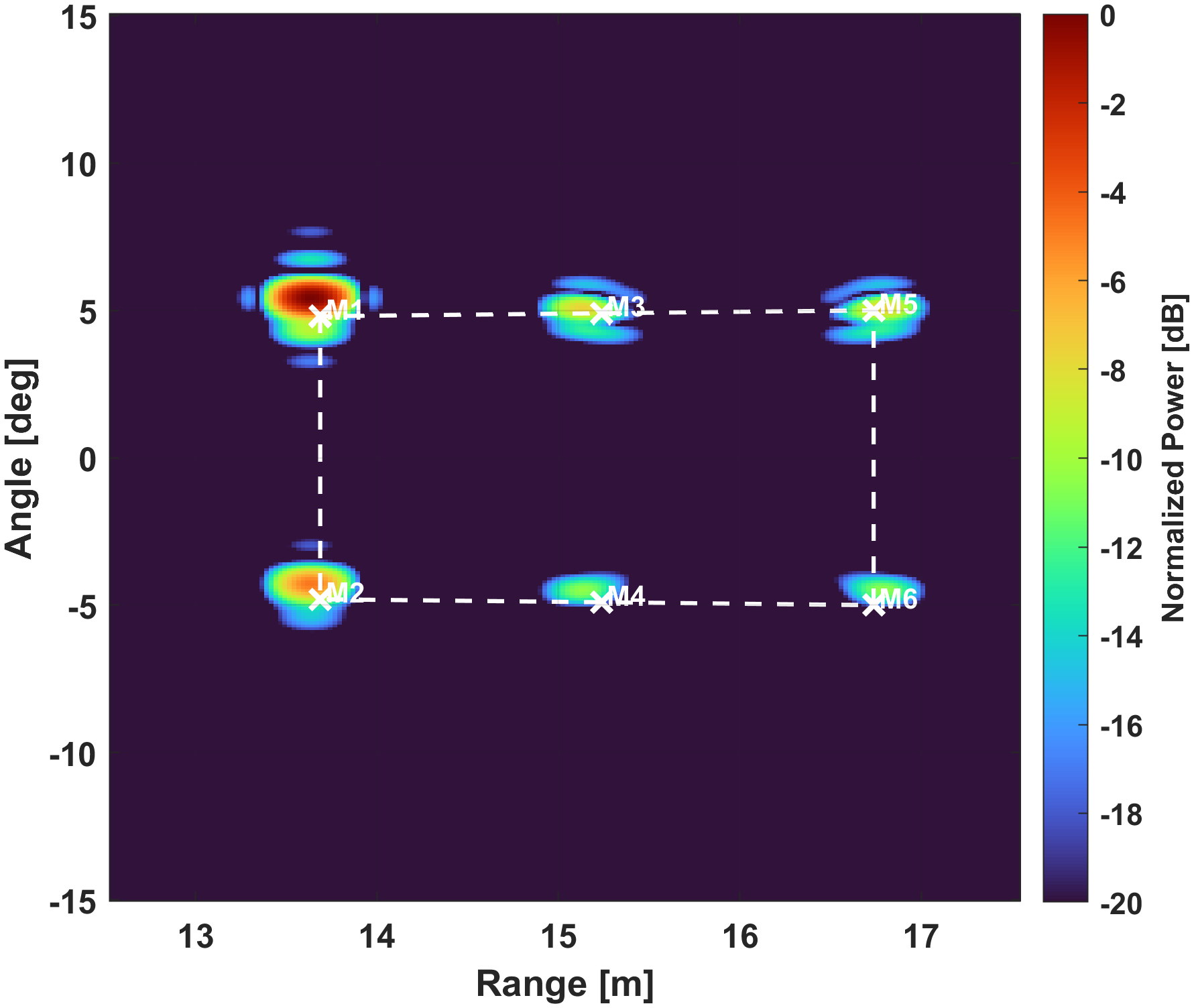}\label{fig:fc}}
\hfill
\subfloat[Sparse Hybrid Beamforming (SC) ISAC-OMP.]{%
\includegraphics[width=0.3\linewidth]{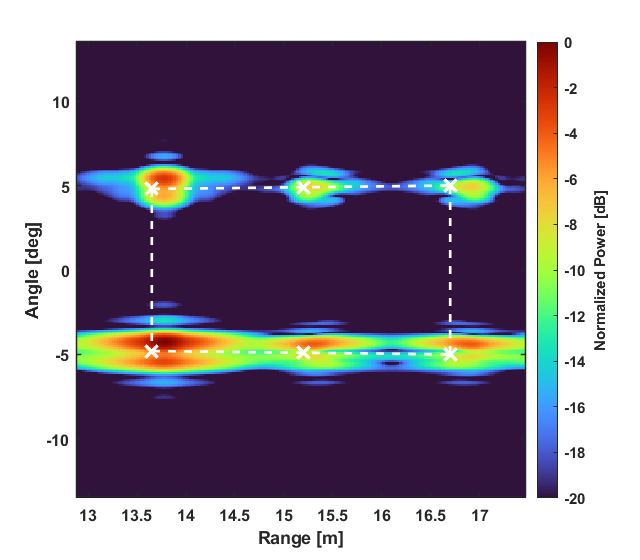}\label{fig:omp}}

\vspace{0.05cm} 

\subfloat[Sparse Hybrid Beamforming (SC) Random-based.]{%
\includegraphics[width=0.3\linewidth]{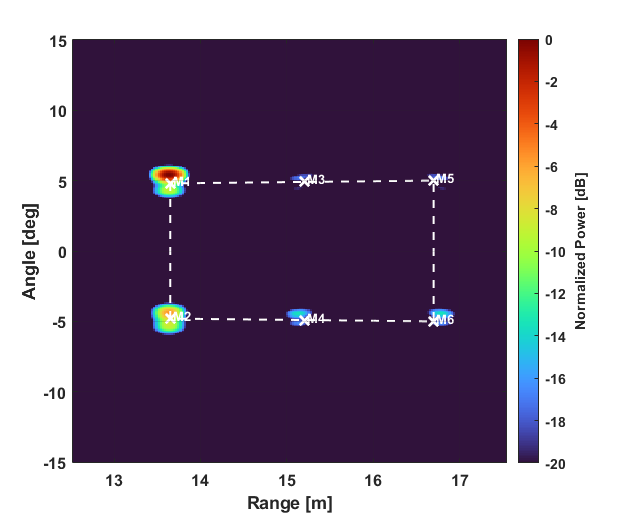}\label{fig:random}}
\hfill 
\subfloat[Sparse Hybrid Beamforming (SC) NA-based.]{%
\includegraphics[width=0.3\linewidth]{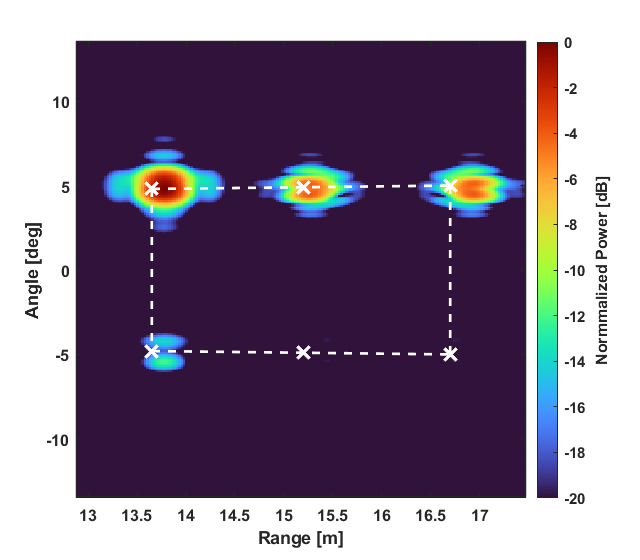}\label{fig:NA}}
\hfill 
\subfloat[Sparse Hybrid Beamforming (SC) Altmin-based.]{%
\includegraphics[width=0.3\linewidth]{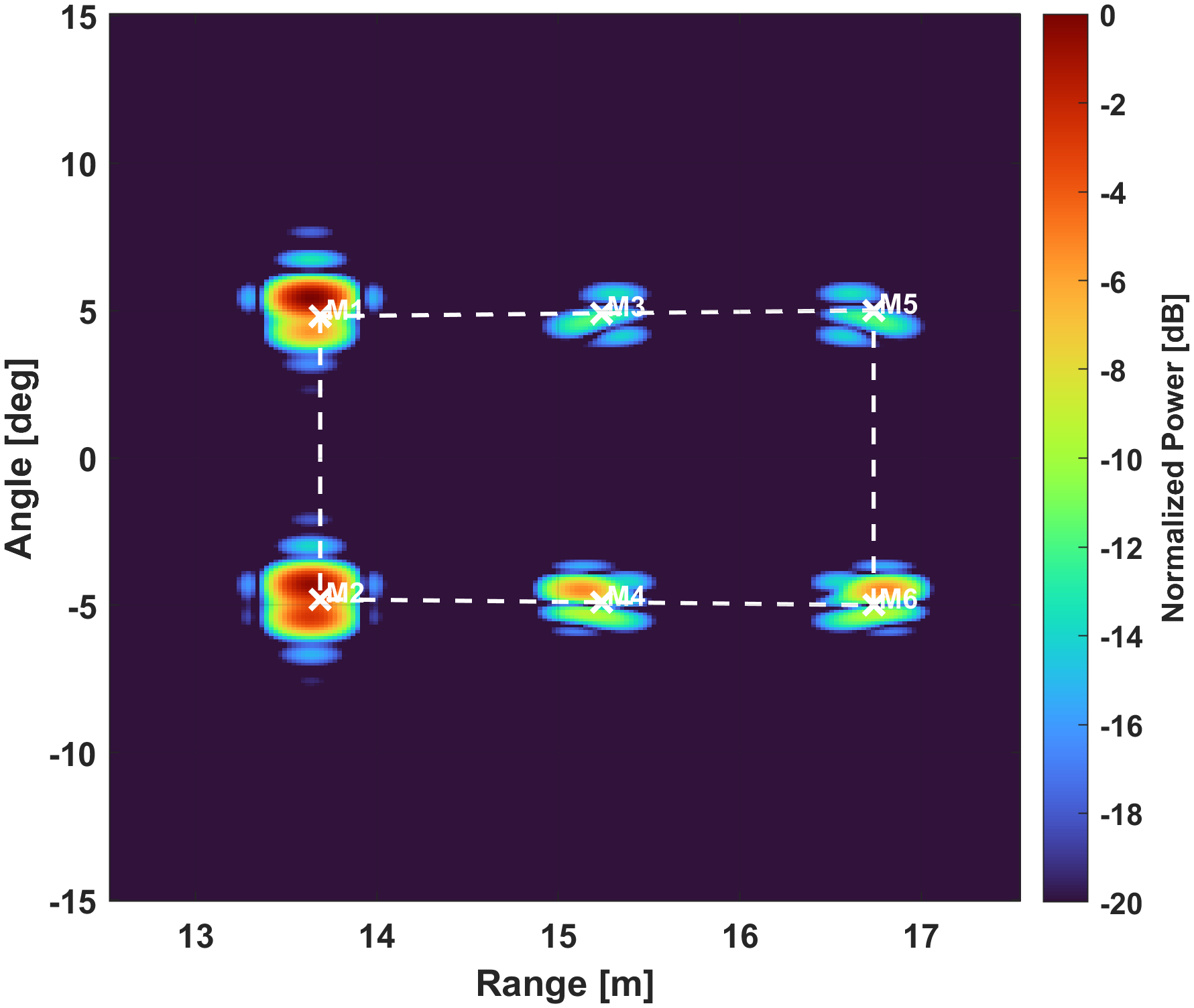}\label{fig:sc}}

\caption{Sensing heatmaps for various beamforming architectures: (a) FD, (b) FC, (c) SC ISAC-OMP-based, (d) SC Random-based, (e) SC NA-based, and (f) SC Altmin-based.}
\label{fig:heatmaps_aligned}
\end{figure*}

We evaluate the sparse-connected ISAC (ISAC-SC) hybrid beamforming scheme in a wideband THz near-field scenario. $N_t = N_r = 128$ antennas with half-wavelength spacing, the carrier frequency is $f_c = 140$~GHz. 
The BS simultaneously serves $K = 4$ 
We consider a single extended target modeled by $M = 6$ dominant scatterers located at positions $\{(r_i,\theta_i)\}_{i=1}^{6}$. \par

The first baseline adopts a random switching strategy, where $\mathbf{P}$ is generated  randomly while preserving the same sparsity percentage as the optimized design. The second baseline is a fixed Nested Array (NA) topology \cite{Pal,10827101,Hunag}: two-level nested structure ($N_1=N_2=12$ per RF chain) 
with the switch matrix $\mathbf P_{\mathrm{NA}}$ fixed by construction so that only Steps~1-2 of Algorithm~\ref{alg:ISAC_SC_altmin_main_refined} are executed. The third baseline is an ISAC-aware OMP (ISAC-OMP) variant that selects antennas using a joint communication-plus-sensing score~\eqref{eq:isac_score}, after which Steps~1-2 are run exactly as in SC-AltMin to obtain a constraint-satisfying precoder.

\begin{equation}
\Delta_i^{\mathrm{ISAC}}
=
\sum_{n=1}^{N_f}
\Big[
(1-\lambda)\,C_i[n]
+
\lambda\,S_i[n]
\Big],
\label{eq:isac_score}
\end{equation}

\subsection{Sensing Performance}
Figure \ref{fig:heatmaps_aligned} presents the range-angle reconstruction of multiple scatters belonging to a single ET under the FD, FC, SC-AltMin, ISAC-OMP, Random-based, and NA-based. This scenario requires preserving the intra-target resolvability of closely spaced scattering centers, which is particularly due to array sparsity. The FD architecture in Fig. \ref{fig:fd} serves as the performance benchmark, exhibiting concentrated main lobes and minimal sidelobe leakage, ensuring perfect intra-target resolvability of all $M$ scattering centers. The FC hybrid structure in Fig. \ref{fig:fc} introduces mild energy spreading, yet maintains distinguishable scatter peaks.\par
 Figure \ref{fig:sc}, the proposed SC-Altmin architecture, despite RF connectivity reduction, preserves peak localization and inter-scatter separation. Although sparsity induces secondary lobes, each scattering center remains distinctly identifiable in both range and angle.\par
As shown in Fig.~\ref{fig:omp}, the SC ISAC-OMP baseline shows incomplete scatterer resolution. Unlike SC-AltMin, it cannot adapt to the evolving residual mismatch between the hybrid precoder and the FD benchmark. The fixed weighting $\lambda=0.5$ biases the selection toward antennas that score well on average. In Fig. \ref{fig:random}, the random-based scheme lacks structured aperture preservation, resulting in degraded spatial diversity. Figure~\ref{fig:NA} illustrates the NA baseline, which maximizes the virtual aperture for generic DOA estimation. However, its switch matrix $\mathbf P_{\mathrm{NA}}$ is determined by the nested geometry and does not adapt to the scatterer distribution of the target under consideration. Consequently, the grating-lobe directions generated by $\mathbf P_{\mathrm{NA}}$ are independent of the scatterer angles within the ROI, some scatterers may not be illuminated, leading to incomplete target reconstruction. To ensure a fair comparison, the digital covariance is optimized using the same Step-1 SDP employed in the SC-AltMin. These results indicate that a sensing-aware selection criterion or a sensing-motivated geometry alone is not sufficient; the switch pattern is adapted to the specific scatterer configuration. SC-AltMin's explicitly minimizes the precoder mismatch against the FD benchmark for the actual target geometry, iteratively refining the switch pattern; neither ISAC-OMP's nor NA's fixed nested geometry provides this target-specific adaptation. \par
Importantly, sparsity-induced grating lobes are not inherently detrimental when jointly optimized with the digital precoder. The proposed covariance-driven AltMin effectively shapes the spatial response to preserve energy concentration at true scatter locations. Thus, the SC architecture achieves FD-comparable extended-target resolvability while substantially reducing RF connectivity.

\subsection{Communication Performance}
\begin{figure}[h]
    \centering
\vspace*{-0.15in}
\includegraphics[width= .75\linewidth]{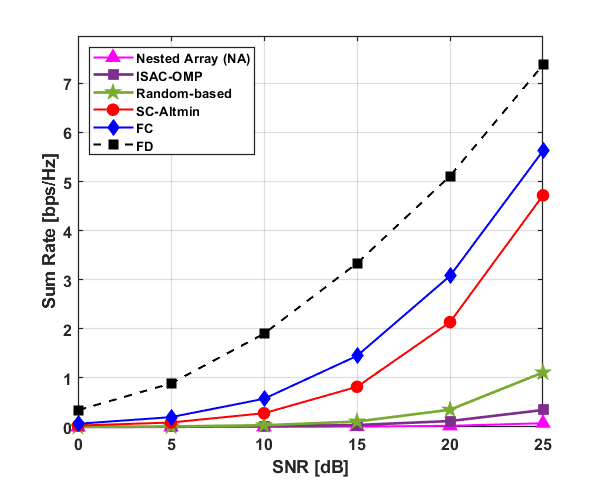}
    \caption{Achievable sum rate versus SNR.}
    \label{comm}
\end{figure}

Figure \ref{comm} shows the achievable sum rate versus SNR for the FD, FC, SC-AltMin, Random-based, ISAC-OMP, and NA-based. As expected, FD achieves the highest rate, followed by FC. Among the sparse-connected schemes, SC-AltMin achieves the highest sum rate, followed by Random-based, then ISAC-OMP, with NA exhibiting the lowest performance. In the THz regime, the channel is low-rank and the number of RF chains exceeds the dominant channel rank. Hence, the rate gap is driven by how accurately each architecture realizes the optimal transmit covariance under structural constraints. The FD design directly aligns the covariance with the dominant channel eigenmodes, while FC approximates it through analog-digital factorization. NA's poor communication performance follows from its design: the nested switch topology is selected purely to maximize array aperture for direction-of-arrival estimation and is independent of the channel directions of the $K=4$ users.
ISAC-OMP's underperformance relative even to the random baseline due to its score. Although includes a communication term $C_i[n]$, this term reflects only single-user channel-gain magnitude rather than the joint multi-user covariance structure; moreover, the fixed weighting $\lambda=0.5$ biases the selection toward antennas that also score well on the sensing term $S_i[n]$, concentrating the resulting switch pattern around a subset of antennas well-aligned with the target but poorly suited to the spatial diversity required to separate $K=4$ users. The random baseline, despite using no selection criterion, samples antenna connections more uniformly across the aperture; since Step~1's digital covariance can exploit this broader, connectivity to approximate the FD benchmark, Random-based achieves better covariance alignment. SC-AltMin avoids this trade-off entirely by directly minimizing the precoder mismatch against the FD benchmark in Step~3, which balances communication and sensing alignment through the actually achieved covariance.

\subsection{Energy Efficiency Performance }
\begin{figure}[H]
    \centering
    \vspace*{-0.15in}
\includegraphics[width=.75\linewidth]{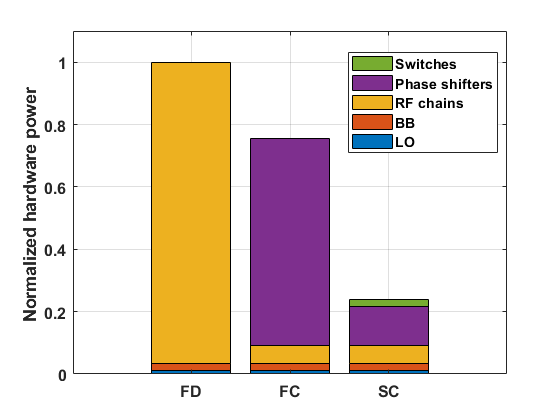}
    \caption{Normalized hardware breakdown power consumption at SNR = 15 dB.}
    \label{EE3}
    \end{figure}

Figure \ref{EE3} illustrates the normalized hardware power consumption of the considered architectures. It is important to note that the SC bar represents all sparse-connected schemes (ISAC-OMP, Random-based, and NA-based), since they share the same RF-chain count and sparsity percentage.\par
In the FD architecture, hardware power is dominated by the RF-chain block due, causing power consumption to scale linearly with array size. In the FC hybrid structure, although the number of RF chains is reduced, the dense phase-shifter network introduces significant analog-domain overhead. In contrast, the SC architecture replaces the fully connected phase-shifter network with low-power switches and reduced RF-chain connectivity. Since switches consume significantly less power, SC achieves the largest hardware reduction. This confirms that analog connectivity rather than digital baseband processing is the dominant energy bottleneck in THz XL-MIMO, and that structured switching provides an effective and scalable solution.

\begin{figure}[htbp]
    \centering
\vspace*{-0.15in}
\includegraphics[width=.75\linewidth]{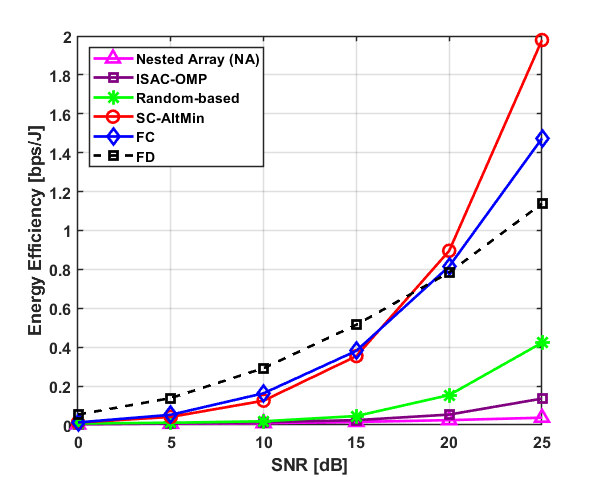}
    \caption{Achievable energy efficiency versus SNR.}
    \label{EE1}
\end{figure}

Figure \ref{EE1} reports the achievable energy efficiency (bit/Joule) versus SNR for all considered schemes. While FD and FC achieve higher spectral efficiency, their large hardware consumption limits energy efficiency. Among the sparse-connected architectures, SC-AltMin attains the highest energy efficiency, followed by Random-based, then ISAC-OMP, with NA exhibiting the lowest performance, directly mirroring the sum-rate ordering of Fig.~\ref{comm}. Since all sparse-connected schemes share identical hardware power, the EE differences stem purely from achievable rate. driven by covariance alignment; SC-AltMin's iteratively-refined switch pattern achieves the best covariance alignment with the FD benchmark, Random's unstructured-but-broad connectivity moderately outperforms ISAC-OMP's narrowly-tuned single-pass selection, and NA's channel-agnostic nested topology yields the worst alignment. The performance gap widens at high SNR, where rate differences dominate since hardware power remains constant. 

\begin{figure*}[h] 
\centering
\subfloat[B = 1.]{%
\includegraphics[width=0.3\linewidth]{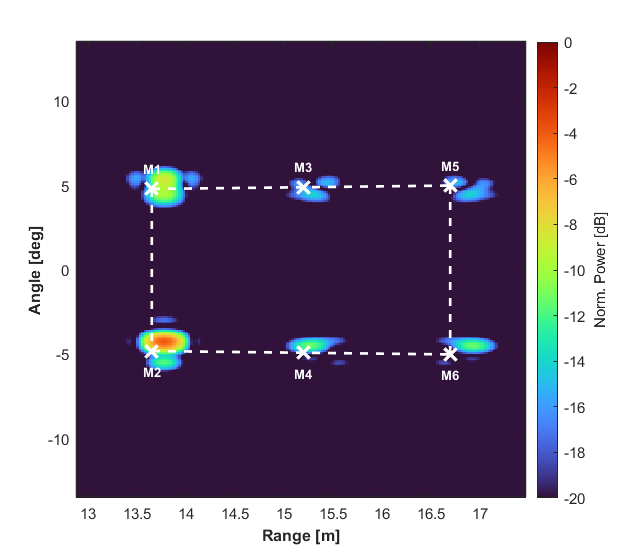}\label{fig:row1_col1}}
\hfill
\subfloat[B = 3.]{%
\includegraphics[width=0.3\linewidth]{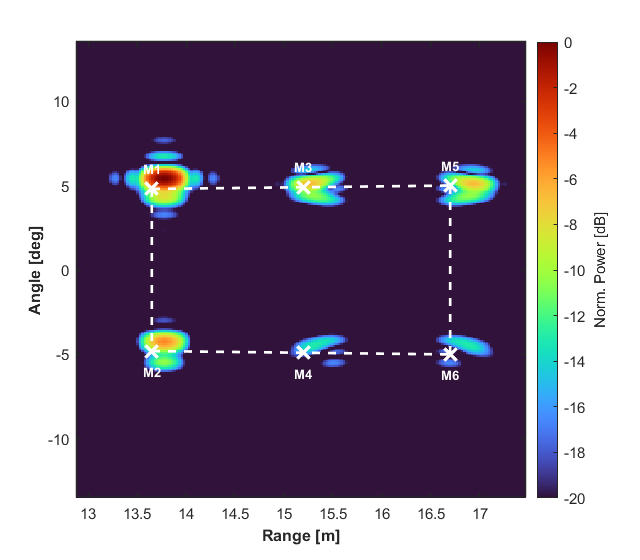}\label{fig:row1_col3}}
\hfill
\subfloat[B = 5.]{%
\includegraphics[width=0.3\linewidth]{Figure/sensing_SC.png}\label{fig:row2_col2}}

\caption{Sensing heatmap for Altmin-SC for different phase quantization levels: (a) B = 1, (b) B = 3, and (c) B = 5.}
\label{fig:main_grid}
\end{figure*}
\subsection{Robustness to Phase Quantization and Antenna Configuration}

\begin{figure}[H]
    \centering
\vspace*{-0.15in}
\includegraphics[width= .75\linewidth]{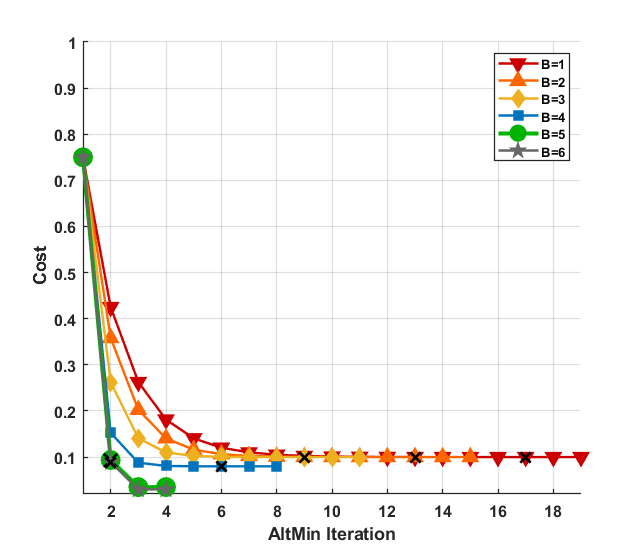}
    \caption{Convergence analysis versus B.}
    \label{conv}
\end{figure}

Figure~\ref{fig:main_grid} shows sensing heatmaps for $B\in\{1,3,5\}$ bits, and Fig. \ref{conv} shows the
AltMin cost versus iteration for $B\in\{1,\ldots,6\}$ bits. In Fig.~\ref{fig:main_grid}, all six
scatterers are correctly localized at the same range-angle positions across all
$B$, confirming that lobe locations are set by the array geometry and switch
pattern, not by phase resolution; only sidelobe sharpness improves with
increasing $B$. In Fig.~\ref{conv}, coarser resolutions
converge more rapidly starting from 17 iterations at $B=1$ , while $B=5$-$6$ converge within $t=2$ iterations, since wider
quantization cells at low $B$ cause phase updates to flip between levels before
settling. All curves reach a comparable final cost, confirming $B$ governs
convergence speed rather than final sensing accuracy. Overall, $B=5$--$6$-bit
phase shifters offer the best trade-off between convergence speed and hardware
complexity, while 1-2 bit shifters eventually reach comparable accuracy at the
cost of more iterations and residual sidelobe leakage.

\begin{figure}[h]
    \centering
\includegraphics[width=.5\linewidth, height=2.75 in, keepaspectratio=false]{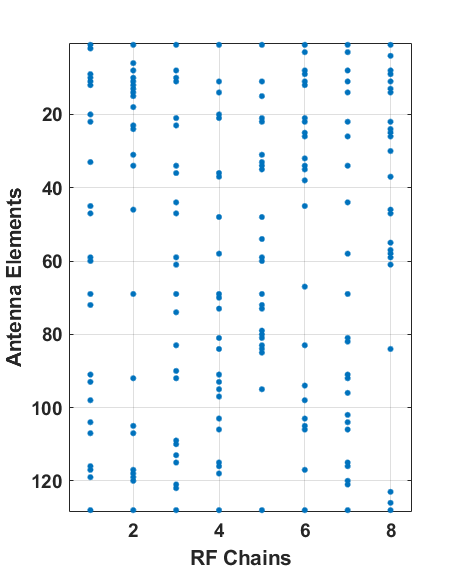}
    \caption{Active antenna link representation.}
    \label{Antenna}
\end{figure}

Finally, Figure \ref{Antenna} shows the optimized antenna RF chain connectivity. A clear structural property emerges: the first and last antenna elements are consistently activated, ensuring that the effective array aperture is maintained. Since angular resolution scales with aperture size, retaining edge elements prevents sparsification from degrading sensing resolution. The remaining active antennas are distributed in a structured manner, reflecting optimization-driven spatial response shaping, to preserves the array’s resolving capability while reducing hardware connectivity. This result confirms that the proposed sparsification strategy is principled: it achieves substantial hardware and energy reduction without compromising  aperture-dependent sensing performance.

\section{Conclusion}
This paper proposed a sparse-connected hybrid beamforming architecture for extended-target THz ISAC that intentionally exploits structured grating lobes to achieve single-shot, full-aperture illumination. By replacing dense analog connectivity with switch-controlled sparsity while preserving aperture, the design enables structured spatial energy shaping without compromising sensing resolution. A covariance-driven alternating-minimization framework jointly optimizes digital precoding, quantized phase shifters, and antenna-RF switching, achieving rapid convergence. Numerical results at 140 GHz demonstrated near fully-digital Cramér-Rao sensing accuracy, competitive communication performance, and substantial reductions in RF-chain utilization and overall energy consumption. These findings confirm that structured sparsity can preserve extended-target resolvability while improving hardware scalability and energy efficiency. Future work will extend the framework to dynamic target tracking and multi targets detection in cross near- and far-fields. 

\section*{Appendix A}
\label{app:convergence}

To study the convergence of the proposed Altmin Algorithm, let
\begin{equation}
f\!\left(\mathbf{P},\mathbf{E},\{\mathbf{F}_{BB}[n]\}\right)
=
\sum_{n=1}^{N_f}
\left\|
\mathbf{W}_{TX}^{FD}[n]
-
(\mathbf{P}\circ\mathbf{E})\mathbf{F}_{BB}[n]
\right\|_F^2
\label{eq:app_obj}
\end{equation}
denote the mismatch objective. The binary switch and quantized phase sets are finite, whereas the covariance set is compact and convex.

\begin{proposition}
\label{prop:app_convergence}
Under Algorithm~3, the sequence $\{f^{(t)}\}$ is monotonically non-increasing and converges to a finite $f^\star\ge 0$. Furthermore, every accumulation point of the iterate sequence is stationary.
\end{proposition}

\textit{Step 1 (Digital covariance update):}
For fixed $\mathbf{P}$ and $\mathbf{E}$, the SDP in (51) is convex. Since the previous iterate remains feasible, $f^{(t,1)} \le f^{(t-1)}$. \label{eq:step1_app}

\textit{Step 2 (Analog phase update):}
Each active phase entry is updated by a least-squares solution followed by projection onto the finite quantized phase, yielding $f^{(t,2)} \le f^{(t,1)}$. \label{eq:step2_app}

\textit{Step 3 (Switch matrix update):}
Algorithm~2 greedily activates links only when the incremental mismatch is nonpositive, yielding $f^{(t)} = f^{(t,3)} \le f^{(t,2)}$. \label{eq:step3_app}

 Thus, $f^{(t)} \le f^{(t-1)}$, $\forall\,t\ge1$. Since $f^{(t)}\ge0$,  the monotone convergence theorem guarantees $f^{(t)} \rightarrow f^\star \ge 0$.


\begin{remark}
Proposition~\ref{prop:app_convergence} guarantees convergence to a stationary point rather than the global optimum, which is standard for 
hybrid beamforming algorithms.
\end{remark}

\begin{remark}
The stabilization in Fig.~8 is consistent with Proposition~\ref{prop:app_convergence}, as the covariance-domain SDP update captures most gain, while the phase and switch updates compensate for residual errors.
\end{remark}

\section*{Appendix B}
\label{app:blkdiag}
For two distinct scatterers $i\neq j$, the cross term in $\mathbf J_{ij}$ involves the product
\begin{equation}
\alpha_i^*[n]\alpha_j[n]
=
|\gamma_i[n]||\gamma_j[n]|\,
e^{j2k_n(r_i-r_j)}\,
e^{j(\phi_j-\phi_i)}.
\label{eq:alphaij_cross}
\end{equation}
The surface phases $\phi_i\sim\mathcal U[0,2\pi)$ are mutually independent, implying $\mathbb E[e^{j(\phi_j-\phi_i)}]=0$ for $i\neq j$, and thus $\mathbb E[\mathbf J_{ij}]=\mathbf 0$. For a fixed realization of the surface phases, the cross-block terms are generally non-zero, and their magnitude is determined by the correlation between respective steering vectors,
\begin{equation}
\rho_{ij}[n]
=
\mathbf a_t^H(r_i,\theta_i;n)\,\mathbf a_t(r_j,\theta_j;n),
\qquad
|[\mathbf J_{ij}]| \propto |\rho_{ij}[n]|^2,
\label{eq:rho_ij_def}
\end{equation}

The block-diagonal approximation~\eqref{eq:J_blkdiag_final_convex} is valid under:
\begin{enumerate} \item independent random surface phases across the scatterers.
\item sufficient angular and/or range separation such that 
$|\rho_{ij}[n]|$ is small;
\item a large array aperture that enhances spatial resolution 
\end{enumerate}

\bibliographystyle{IEEEtran}
\bibliography{references}

@ARTICLE{10439221,
  author={Han, Chong and Wu, Yongzhi and Chen, Zhi and Chen, Yi and Wang, Guangjian},
  journal={IEEE Commun. Mag.}, 
  title={{THz} {ISAC}: A Physical-Layer Perspective of Terahertz Integrated Sensing and Communication}, 
  year={2024},
month={Feb.},
  volume={62},
  number={2},
  pages={102-108},
  doi={10.1109/MCOM.001.2200404}}

@ARTICLE{11060909,
  author={Zhang, Haobo and Huang, Xiangchao and Guo, Xiuzhen and He, Shibo and Gu, Chaojie and Shu, Yuanchao and Chen, Jiming},
  journal={IEEE Trans. Netw. Sci. Eng.}, 
  title={Terahertz Sensing, Communication, and Networking: A Survey}, 
  year={2026},
month={Jul.},
  volume={13},
  number={},
  pages={501-521},
  doi={10.1109/TNSE.2025.3584922}}

@ARTICLE{10422712,
  author={Xue, Qing and Ji, Chengwang and Ma, Shaodan and Guo, Jiajia and Xu, Yongjun and Chen, Qianbin and Zhang, Wei},
  journal={IEEE Commun. Surveys Tuts.}, 
  title={A Survey of Beam Management for mmWave and {THz} Communications Towards {6G}}, 
  year={2024},
month={Feb.},
  volume={26},
  number={3},
  pages={1520-1559},
  doi={10.1109/COMST.2024.3361991}}

@article{elbir2024terahertz,
  title={Terahertz-band integrated sensing and communications: Challenges and opportunities},
  author={Elbir, Ahmet M and Mishra, Kumar Vijay and Chatzinotas, Symeon and Bennis, Mehdi},
  journal={IEEE Aerosp. Electron. Syst. Mag},
  year={2024},
  month={Dec.},
  volume={39},
  number={12},
  pages={38-49},
  publisher={IEEE},
  doi={10.1109/MAES.2024.3476228}
}

@article{rafique2025terahertz,
  title={Terahertz-Band Channel: An Enabler for Futuristic Wireless Sensing and Communication},
  author={Rafique, Saira and Abusanad, Nusaibah and Arslan, H{\"u}seyin},
  journal={IEEE Veh. Technol. Mag.},
  year={2026},
  month= {Mar.},
  volume={21},
  number={1},
  pages={60-70},
  doi= {10.1109/MVT.2025.3579576},
  publisher={IEEE}
}

@ARTICLE{11153494,
  author={Zhang, Fan and Mao, Tianqi and Li, Mingkun and Hua, Meng and Chen, Jinshu and Masouros, Christos and Wang, Zhaocheng},
  journal={IEEE Network}, 
  title={Near-Field {ISAC} for {THz} Wireless Systems}, 
  year={2025},
month={Nov.},
  volume={39},
  number={6},
  pages={54-61},
  doi={10.1109/MNET.2025.3598827}}

@article{shen2025hybrid,
  title={Hybrid Beamforming with Widely-spaced-array for Multi-user Cross-Near-and-Far-Field Communications},
  author={Shen, Heyin and Chen, Yuhang and Han, Chong and Yuan, Jinhong},
  journal={IEEE Trans. Commun.},
  year={2025},
month={Sep.},
  volume={73},
  number={9},
  pages={7858-7873},
  doi= {10.1109/TCOMM.2025.3545592},
  publisher={IEEE}
}

@ARTICLE{10475421,
  author={Elbir, Ahmet M. and Abdallah, Asmaa and Celik, Abdulkadir and Eltawil, Ahmed M.},
  journal={IEEE J. Sel. Top. Signal Process.}, 
  title={Antenna Selection With Beam Squint Compensation for Integrated Sensing and Communications}, 
  year={2024},
 month={Jul.},
  volume={18},
  number={5},
  pages={857-870},
  doi={10.1109/JSTSP.2024.3378909}}

@article{yang2025sensing,
  title={Sensing-Centric Energy-Efficient Hybrid Beamforming Design for Terahertz Integrated Sensing and Communications},
  author={Yang, Kai and Xu, Jin and Tao, Xiaofeng and Sun, Mengying and Wang, Bing and Wu, Huici},
  journal={IEEE Trans. Veh. Technol},
  year={2025},
  month={Sep.},
   volume={74},
  number={9},
  pages={14887-14892},
  doi= {10.1109/TVT.2025.3561170},
  publisher={IEEE}
}

@inproceedings{elbir2023near,
  title        = {Near-Field Hybrid Beamforming for Terahertz-Band Integrated Sensing and Communications},
  author       = {Elbir, Ahmet M. and Celik, Abdulkadir and Eltawil, Ahmed M.},
  booktitle    = {Proceedings of the 2023 IEEE Globecom Workshops (GC Wkshps)},
  pages        = {1117--1122},
  year         = {2023},
  month={Dec.},
  organization = {IEEE},
  address      = {Kuala Lumpur, Malaysia},
  doi= {10.1109/GCWkshps58843.2023.10464709},
}

@article{yan2022energy,
  title={Energy-efficient dynamic-subarray with fixed true-time-delay design for terahertz wideband hybrid beamforming},
  author={Yan, Longfei and Han, Chong and Yuan, Jinhong},
  journal={IEEE J. Sel. Areas Commun.},
  volume={40},
  number={10},
  pages={2840--2854},
  year={2022},
month={Oct.},
doi= {10.1109/JSAC.2022.3196090},
  publisher={IEEE}
}

@inproceedings{wang2024cramer,
  title        = {Cram{\'e}r-Rao Bound Analysis and Beamforming Design for {3D} Extended Target in {ISAC}},
  author       = {Wang, Yiqiu and Tao, Meixia and Sun, Shu and Cao, Wei},
  booktitle    = {Proceedings of the 2024 IEEE Global Communications Conference (GLOBECOM)},
  pages        = {5344--5349},
  year         = {2024},
  organization = {IEEE},
doi= {10.1109/GLOBECOM52923.2024.10901196},
  address      = {Cape Town, South Africa}
}

@article{du2022integrated,
  title={Integrated sensing and communications for {V2I} networks: Dynamic predictive beamforming for extended vehicle targets},
  author={Du, Zhen and Liu, Fan and Yuan, Weijie and Masouros, Christos and Zhang, Zenghui and Xia, Shuqiang and Caire, Giuseppe},
  journal={IEEE Trans. Wireless Commun.},
  volume={22},
  number={6},
  pages={3612--3627},
  year={2023},
month={Jun.},
 doi={10.1109/TWC.2022.3219890},
  publisher={IEEE}
}

@inproceedings{wang2023beamforming,
  title        = {Beamforming Design for Integrated Sensing and Communication with Extended Target},
  author       = {Wang, Yiqiu and Tao, Meixia and Sun, Shu},
  booktitle    = {Proceedings of the 2023 IEEE Globecom Workshops (GC Wkshps)},
  pages        = {1392--1397},
  year         = {2023},
  organization = {IEEE},
  doi={10.1109/GCWkshps58843.2023.10465036},
  address      = {Kuala Lumpur, Malaysia}
}

@article{yao2025hybrid,
  author={Yao, Yu and Zhang, Junhao and Miao, Pu and Zhang, Long and Chen, Gaojie and Shu, Feng and Wong, Kai-Kit},
  journal={IEEE Trans. Commun.}, 
  title={Hybrid {RIS}-Enhanced {ISAC} Secure Systems: Joint Optimization in the Presence of an Extended Target}, 
  year={2025},
  month={Dec.},
  volume={73},
  number={12},
  pages={15688-15704},
  doi={10.1109/TCOMM.2025.3610218}}

@inproceedings{wang2025deep,
author    = {Wang, Y. and Tao, M. and Sun, S.},
  booktitle = {Proc. IEEE Int. Conf. Commun. Workshops (ICC Wkshps)},
  title     = {Deep Learning-Based Extended Target Tracking in {ISAC} Systems},
  year      = {2025},
  pages     = {678--683},
  doi       = {10.1109/ICCWorkshops67674.2025.11162407},
  address   = {Montreal, QC, Canada},
  month     = {Jun.}
}

@article{zhao2024joint,
  title={Joint target localization and data detection in bistatic {ISAC} networks},
  author={Zhao, Na and Chang, Qing and Shen, Xiao and Wang, Yunlong and Shen, Yuan},
  journal={IEEE Trans. Commun.},
  year={2025},
month={May.},
 volume={73},
  number={5},
  pages={3531-3546},
  doi= {10.1109/TCOMM.2024.3481046},
  publisher={IEEE}
}

@article{wang2023propagation,
  title={Propagation characteristics and modeling of monostatic RCS scattering from indoor building material above {215-GHz} frequency bands},
  author={Wang, Yang and Li, Chuan and Liao, Xi and Yu, Ziming and Wang, Guangjian and Li, Xianjin and Zhang, Jie},
  journal={IEEE Wirel. Commun. Lett.},
  volume={13},
  number={3},
  pages={642--646},
  year={2024},
month={Mar.},
 doi={10.1109/LWC.2023.3338211},
  publisher={IEEE}
}

@article{rafique2025rough,
  title={A Rough Surface-Aided Non-Line-of-Sight Joint Sensing and Communication System for {THz} Band},
  author={Rafique, Saira and Arslan, H{\"U}Seyin},
  journal={IEEE Open J. Commun. Soc.},
  year={2025},
month={Mar.},
volume={6},
  number={},
  pages={2239-2255},
   doi= {10.1109/OJCOMS.2025.3552635},
  publisher={IEEE}
}

@article{li2025sparse,
  author={Li, Xinrui and Min, Hongqi and Zeng, Yong and Jin, Shi and Dai, Linglong and Yuan, Yifei and Zhang, Rui},
  journal={IEEE Wireless Commun.}, 
  title={Sparse {MIMO} for {ISAC}: New Opportunities and Challenges}, 
  year={2025},
  volume={32},
  number={4},
  month={Aug.},
  pages={170-178},
  doi={10.1109/MWC.001.2400201}}

@inproceedings{evmorfos2024generative,
author    = {Evmorfos, Spilios and Petropulu, Athina P.},
  title     = {Generative AI for Sparse Antenna Array Design in ISAC Systems},
  booktitle = {2024 IEEE 25th International Workshop on Signal Processing Advances in Wireless Communications (SPAWC)},
  year      = {2024},
  pages     = {306--310},
  address   = {Lucca, Italy},
  doi       = {10.1109/SPAWC60668.2024.10694360},
  publisher = {IEEE}
}

@article{liu2024doa,
  title={{DOA} estimation-oriented joint array partitioning and beamforming designs for ISAC systems},
  author={Liu, Rang and Li, Ming and Liu, Qian and Swindlehurst, A Lee},
  journal={IEEE Trans. Wireless Commun.},
  year={2025},
month={Mar.},
  volume={24},
  number={3},
  pages={ 2052-2066},
doi       = {10.1109/TWC.2024.3516037},
  publisher={IEEE}
}

@inproceedings{abusanad2024joint,
  title        = {Joint Imaging and Communication Using Sparse Antenna Arrays},
  author       = {Abusanad, Nusaibah A. and Ayasrah, Mus' ab and Arslan, H{\"u}seyin},
  booktitle    = {Proceedings of the 2024 IEEE Wireless Communications and Networking Conference (WCNC)},
  pages        = {1--6},
  year         = {2024},
  organization = {IEEE},
  doi={ 10.1109/WCNC57260.2024.10571314},
  address      = {Dubai, United Arab Emirates}
}

@article{dorvash2025virtual,
author    = {Dorvash, Masoud and Naghdi, Mohammad Javad and Zahabi, Sayed Jalal and Modarres-Hashemi, Mahmoud and Alaee-Kerahroodi, Mohammad and Lang, Oliver and Feger, Reinhard},
  title     = {Virtual Array Design via Hole-Free Sparse Array With Reduced Mutual Coupling Effect},
  journal   = {IEEE Trans. Aerosp. Electron. Syst.},
  year      = {2025},
  month={Dec.},
  volume    = {61},
  number    = {6},
  pages     = {18702--18718},
  doi       = {10.1109/TAES.2025.3617024},
  publisher = {IEEE}
}

@article{wang2024wideband,
  title={Wideband near-field integrated sensing and communication with sparse transceiver design},
  author={Wang, Xiangrong and Zhai, Weitong and Wang, Xianghua and Amin, Moeness G and Cai, Kaiquan},
  journal={IEEE J. Sel. Topics Signal Process.},
  volume={18},
  number={4},
  pages={662--677},
  year={2024},
 month={May.},
 doi= {10.1109/JSTSP.2024.3394970},
  publisher={IEEE}
}

@ARTICLE{11220947,
  author={Alshorman, Nusaibah A. and Aïssa, Sonia and Arslan, Hüseyin},
  journal={IEEE Internet Things J.}, 
  title={Integrated Sensing and Communication Beamforming Design with Target Model Aware Antenna Selection}, 
  year={2026},
month={Jan.},
  volume={13},
  number={1},
  pages={963-977},
  doi={10.1109/JIOT.2025.3626821}}

@ARTICLE{11159279,
  author={Zhang, Yuxiang and Zhang, Jianhua and Gong, Huiwen and Hu, Xidong and Zhang, Jiwei and Xing, Hongbo and Luo, Shilin and Xiong, Yifeng and Yu, Li and Yuan, Zhiqiang and Liu, Guangyi and Jiang, Tao},
  journal={IEEE J. Sel. Areas Commun.}, 
  title={A Unified {RCS} Modeling of Typical Targets for {3GPP} {ISAC} Channel Standardization and Experimental Analysis}, 
  year={2025},
  month={Sep.},
  volume={44},
  number={},
  pages={702-716},
  doi={10.1109/JSAC.2025.3608732}}

@ARTICLE{liu2025nf,
  author={Liu, Yuanwei and Xu, Jiaqi and Wang, Zhaolin and Mu, Xidong and Hanzo, Lajos},
  journal={IEEE Wireless Commun.}, 
  title={Near-field Communications: What Will Be Different?}, 
  year={2025},
  month={Apr.},
  volume={32},
  number={2},
  pages={262-270},
  doi={10.1109/MWC.001.2300588}}

@ARTICLE{zhang2025nf,
  author={Cui, Mingyao and Wu, Zidong and Lu, Yu and Wei, Xiuhong and Dai, Linglong},
  journal={IEEE Commun. Mag.}, 
  title={Near-Field {MIMO} Communications for {6G}: Fundamentals, Challenges, Potentials, and Future Directions}, 
  year={2023},
  month={Jan.},
  volume={61},
  number={1},
  pages={40-46},
  doi={10.1109/MCOM.004.2200136}}

@article{garcia2017reduced,
  title={Reduced switching connectivity for large scale antenna selection},
  author={Garcia-Rodriguez, Adrian and Masouros, Christos and Rulikowski, Pawel},
  journal={IEEE Trans. Commun.},
  volume={65},
  number={5},
  pages={2250--2263},
  year={2017},
  month={May.},
  doi={10.1109/TCOMM.2017.2669030},
  publisher={IEEE}
}

@article{sobolewski2022state,
  title={State of the art sub-terahertz switching solutions},
  author={Sobolewski, Jakub and Yashchyshyn, Yevhen},
  journal={IEEE Access},
  volume={10},
  pages={12983--12999},
  year={2022},
  month={Jan.},
  doi={10.1109/ACCESS.2022.3147019},
  publisher={IEEE}
}

@article{ziegahn2025full,
  title={Full Duplex Transmit and Receive Beamforming With Block-Sparse Antenna Selection for Multi-User Massive{ MIMO}},
  author={Ziegahn, Richard and Le-Ngoc, Tho},
  journal={IEEE Open J. Commun. Soc.},
  volume={6},
  pages={10287--10306},
  year={2025},
  month={Dec.},
  doi={10.1109/OJCOMS.2025.3640594},
  publisher={IEEE}
}

@ARTICLE{11039147,
  author={Zhou, Cong and You, Changsheng and Zhang, Haodong and Chen, Li and Shi, Shuo},
  journal={IEEE Trans. Wireless Commun.}, 
  title={Sparse Array Enabled Near-Field Communications: Beam Pattern Analysis and Hybrid Beamforming Design}, 
  year={2025},
  month={Dec.},
  volume={24},
  number={12},
  pages={10261-10277},
  doi={10.1109/TWC.2025.3578561}}

@article{hu2023,
  author    = {Hu, L. and Wu, J. and Zhang, L.},
  journal   = {Remote Sensing},
  title     = {Efficient transceiving search scheme and implementation method for collocated distributed coherent aperture radar via grating lobes exploitation},
  year      = {2023},
  month     = {Apr.},
  volume    = {15},
  number    = {9},
  pages     = {2262},
  doi       = {10.3390/rs15092262},
  publisher = {MDPI}
}

@article{xu2023,
  author    = {Gao, F. and Xu, L. and Ma, S.},
  journal   = {IEEE Trans. Commun.},
  title     = {Integrated Sensing and Communications With Joint Beam-Squint and Beam-Split for {mmWave/THz} Massive {MIMO}},
  year      = {2023},
  month     = {May.},
  volume    = {71},
  number    = {5},
  pages     = {2963--2976},
  doi       = {10.1109/TCOMM.2023.3245657},
  publisher = {IEEE}
}

@article{Pal,
  author    = {Pal, Piya and Vaidyanathan, P. P.},
  journal   = {IEEE Trans. Signal Process.}, 
  title     = {Nested Arrays: A Novel Approach to Array Processing With Enhanced Degrees of Freedom}, 
  year      = {2010},
  volume    = {58},
  number    = {8},
  pages     = {4167-4181},
  doi       = {10.1109/TSP.2010.2049264},
  month     = aug
}

@inproceedings{10827101,
  author    = {Min, Hang and Feng, Chi and Li, Rong and Zeng, Yong},
  booktitle = {Proc. IEEE Int. Conf. Wireless Commun. Signal Process. (WCSP)}, 
  title     = {Integrated Sensing and Communication with Nested Array: Beam Pattern and Performance Analysis}, 
  year      = {2024},
  volume    = {},
  number    = {},
  pages     = {764-769},
  doi       = {10.1109/WCSP62071.2024.10827101},
  month     = oct
}

@inproceedings{Hunag,
  author    = {Huang, Mengting and Yin, Meng and Xu, Huaiyuan},
  booktitle = {Proc. IEEE Wireless Commun. Networking Conf. (WCNC)}, 
  title     = {Analysis on {UAV} Sensing with Sparse Array in {ISAC} System}, 
  year      = {2025},
  volume    = {},
  number    = {},
  pages     = {1-6},
  doi       = {10.1109/WCNC61545.2025.10978166},
  month     = mar
}

@article{zhang2025,
  author    = {Zhang, F. and Mao, T. and Li, M. and Hua, M. and Chen, J. and Masouros, C. and Wang, Z.},
  journal   = {IEEE Netw.},
  title     = {Near-Field {ISAC} for {THz} Wireless Systems},
  year      = {2025},
  month     = {Nov.},
  volume    = {39},
  number    = {6},
  pages     = {54--61},
  doi       = {10.1109/MNET.2025.3598827},
  publisher = {IEEE}
}
\end{document}